\documentclass[aps,twocolumn,showpacs,superscriptaddress,longbibliography]{revtex4-1} 
\usepackage[english]{babel}
\usepackage{amsmath,amssymb,amsfonts,bm,graphicx,verbatim,mathrsfs,units,color}
\usepackage[colorlinks=true,citecolor=blue]{hyperref} 

\usepackage{xcolor}

\newcommand{\kb}{k_\mathrm{B}}

\newcommand{\T }{\mathsf{T}}
\DeclareMathOperator{\Tr}{Tr}

\begin{document}
\title{Extreme-Value Statistics of Stochastic Transport Processes: \\ Applications to Molecular Motors and Sports}

\author{Alexandre Guillet}
 \affiliation{Univ. Bordeaux, CNRS, LOMA, UMR 5798, F-33400 Talence, France}%Universit\'e Bordeaux, Laboratoire Ondes et Mati\`ere d'Aquitaine, 33405 Talence, France}

\author{\'Edgar Rold\'an}
 \affiliation{ICTP - The Abdus Salam International Centre for Theoretical Physics, Strada Costiera 11, 34151, Trieste, Italy}

\author{Frank J\"ulicher}
\affiliation{Max Planck Institute for the Physics of Complex Systems, N\"othnizer Strasse 38, 01187 Dresden, Germany}

\parskip 0.9mm
\def\d{{\rm d}}
\def\Ps{{\mathsf{P}_{\scriptscriptstyle \hspace{-0.3mm} s}}}
\def\MF{{\mbox{\tiny \rm \hspace{-0.3mm} MF}}}

\begin{abstract}
We derive exact expressions for the finite-time statistics of extrema (maximum and minimum) of the spatial displacement and the fluctuating entropy flow of biased random walks. Our approach captures key features of extreme events in 
molecular motor motion along linear filaments.  Our results generalize the infimum law for entropy production and reveal a symmetry of the distribution of its maxima and minima. 
We also show that the relaxation spectrum of the full generating function, and hence of any moment,  of the   finite-time  extrema distributions can be written in terms of the Mar{\v{c}}enko-Pastur distribution of random-matrix  theory. Using this result, we obtain estimates for the extreme-value statistics of stochastic transport from the eigenvalue distributions of suitable Wishart and Laguerre random matrices.
We confirm our results  by  numerical simulations of stochastic models of molecular motors and discuss as illustrative example 
our theory in the context of sports.
\end{abstract}

\maketitle

\section{Introduction}

Life is a non-equilibrium phenomenon characterised by  fluxes of energy and matter at different scales. At the molecular level,  molecular motors play a key role for the generation 
of movements and forces in cells. Examples are vesicle transport, muscle contraction, cell division and 
cell locomotion~\cite{julicher1997modeling,howard2001mechanics}.
A molecular motor consumes a chemical fuel, adenosine triphosphate (ATP),
that is hydrolysed to  adenosine diphosphate (ADP) and inorganic phosphate. The chemical energy
of this reaction is transduced to generate spontaneous movements and mechanical work.
 Single-molecule 
experiments have revealed that the activity of single or a few molecular motors displays strong  fluctuations~\cite{noji1997direct,bustamante2000grabbing,schnitzer2000force,ishii2000single,helenius2006depolymerizing,ritort2006single,vogel2009self,aathavan2009substrate,ananthanarayanan2013dynein,fakhri2014high,naganathan2016actomyosin} which can be captured by the theory of stochastic processes~\cite{ndlec1997self,keller2000mechanochemistry,qian2005cycle,klumpp2005cooperative,campas2006collective,kolomeisky2007molecular,brangwynne2008cytoplasmic,gaspard2016template}.

An important question is to understand general features
and universal properties that govern the statistics of fluctuations of stochastic transport
processes that include the motion of molecular motors.
Universal relations for the fixed-time statistics of time-integrated currents, such as the distance traveled
and the work performed, have been investigated  in the framework of non-equilibrium stochastic thermodynamics~\cite{sekimoto1997kinetic,lau2007nonequilibrium,chemla2008exact,toyabe2010nonequilibrium}. These results provide e.g. universal bounds for the  efficiency of molecular motors given by the ratio between the mechanical power    and the chemical  power put in the motor~\cite{pietzonka2016universal}. Timing statistics of enzymatic reactions, such as those powering the motion of molecular motors, have been discussed within the framework of Kramers theory~\cite{hanggi1990reaction}. Recent theory and experiments in Kinesin  have revealed symmetry relations between forward and backward cycle-time distributions of enzymatic reactions~\cite{qian2006generalized,nishiyama2002chemomechanical,carter2005mechanics}. Related results have been derived in the context of waiting times of active molecular processes~\cite{neri2017statistics} and transition-path times in folding transitions of DNA hairpins~\cite{gladrow2019experimental}.

When discussing dynamic processes, it is also sufficient to study averages and small fluctuations. However, rare events and large fluctuations play an important role when resilience and reliability of a system are investigated. In this context, the statistics of extreme values and of extreme excursions from the average play an important role, as has been discussed in fields ranging from statistical mechanics to climate~\cite{cheng2014non,diffenbaugh2005fine} and finance~\cite{rocco2014extreme,brodin2014extreme,novak2011extreme}. Extreme events are also important in biophysics and are key to understand the robustness of biological processes. Illustrating examples are microtubule catastrophes or a sperm winning a race against a billion competitors. Here we consider the extreme value statistics of transport processes.
In spite of the significant progress in research on steady-state currents and time fluctuations,  little is known yet  about  extreme-value statistics of stochastic 
processes such as active biomolecular processes.  Extreme-value theory has provided useful insights for e.g.   long-range correlations of DNA sequences~\cite{peng1992long,arneodo1995characterizing} and    DNA replication statistics in frogs' embryonic cells~\cite{bechhoefer2007xenopus}. 
However, extreme-value statistics of molecular motors have not been discussed so far. 
For example, what is the maximal excursion  of a stochastic 
motor against or in the average direction of its motion within a given time?  How long does it take a motor to reach its maximum excursion against the chemical bias? What is the entropy production associated with an extreme fluctuation of a molecular motor?

In this article,  we provide  novel insights on the aforementioned questions by deriving  exact results for  
extreme-value statistics of simple models of stochastic transport given in Eqs.~(\ref{eq:PSmin}-\ref{eq:Smin}) and~(\ref{eq:integral}-\ref{SminRMT}). We discuss the statistics of the maximum and minimum excursion (with respect to its initial location) and 
the associated extremal entropy changes. Moreover, we investigate the  timescales associated with those extrema, 
combining concepts from stochastic thermodynamics, random walks and random-matrix theory. As we show below, 
our results provide  insights on extreme-value statistics beyond recently
derived inequalities for the finite-time infimum statistics of entropy production~\cite{chetrite2011two,neri2017statistics}, and  relate to  
record statistics of correlated stochastic processes~\cite{krug2007records,majumdar2008universal,wergen2011record,godreche2014universal,harris2015random,benichou2016temporal,hartich2019extreme,sabhapandit2019extremes}.

The article is organized as follows: Sec.~\ref{sec2} describes the stochastic model  of stochastic transport used in this paper and provides exact extreme-value statistics for one dimensional (1D) biased random walks. Sec.~\ref{sec3} discusses the connection between extrema of 1D biased random walks and random matrix theory. In Sec.~\ref{sec4} we apply our theory to two-dimensional stochastic models of molecular motors and
discuss its implications using sports as illustrative example. Sec.~\ref{sec5} concludes the paper. Details on the mathematical derivations and numerical simulations are provided in the Appendices.
 
\begin{figure}[ht]
\includegraphics[width=7.5cm]{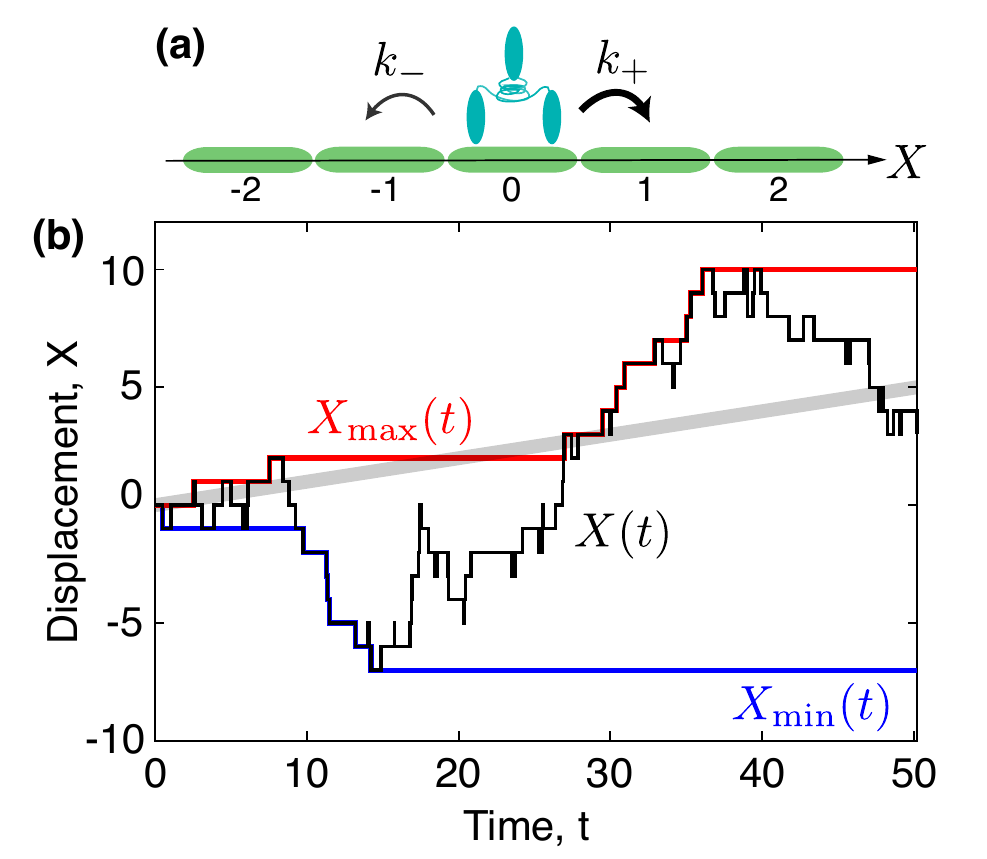}
\caption{(a) Sketch of a one-dimensional (1D) biased random walk with hopping rates $k_+$ and $k_-$ with displacement $X$. (b) Example of a trajectory $X(t)$ (black),  its  maximum $X_{\max}(t)$ (red),   minimum $X_{\min}(t)$ (blue) and its average  over many realizations $\langle X(t)\rangle$ (thick gray) as a function of time $t$. The trajectories are obtained from a numerical simulation of a 1D biased random walk with hopping rates $k_{+}=1.05$ and $k_- =  0.95$ in the positive and negative direction, respectively. The entropy production along the trajectory $X(t)$ is $S(t) = AX(t)$, with $A = \ln (k_+/k_-) = 0.1$. }
\label{fig:1} 
\end{figure}

  \section{Extreme-value statistics of 1D biased random walks}
  \label{sec2}
Many nonequilibrium phenomena at mesoscales can be described at a coarse-grained level as a continuous-time Markov jump process between discrete states $x,y,z$ etc, with exponential waiting times. The transition rate from  $x$ to  $y$ can be written as~\cite{van1992stochastic,bressloff2014stochastic,baiesi2018life}
\begin{equation}
k(x, y) = \nu(x,y)e^{A(x,y)/2}\quad,
\label{eq:1}
\end{equation}
with $\nu(x,y)=\nu(y,x)$ symmetric and $A(x,y)=-A(y,x)$ antisymmetric with respect to the exchange $x\to y$.
If local detailed balance holds $A(x,y)=\beta [ W(x)-W(y)]$ with $W(x)$ the potential energy of state $x$,  $\beta =(k_{\rm B}T)^{-1}$ with $k_{\rm B}$ Boltzmann's constant and $T$ the temperature of a thermostat.
 First,  we  consider the simple case of a 1D biased random walk  on a line with discrete states  denoted by the integer $x \in \mathbb{Z}$ (see Fig.~\ref{fig:1}a for an illustration).
The forward 
 and backward 
jump rates are given by $k_{\pm} \equiv k(x, x\pm 1) = \nu e^{\pm A/2}$. 
Here $\nu>0$ is  a rate and $A>0$ the affinity, which satisfy
 \begin{equation}
 \nu = \sqrt{k_+ k_-}, \quad A=\ln (k_+/k_-)\quad.
 \label{eq:2}
 \end{equation}
   This biased random walk describes e.g. the motion of a molecular motor along a periodic track fuelled by ATP~\cite{kolomeisky2007molecular}. 
   In the simplest case,  
   $A=\beta\Delta \mu$, with $\Delta \mu=\mu_{\rm ATP}-(\mu_{\rm ADP}+\mu_P)$ the chemical potential difference of ATP hydrolysis, often of the order of $20 k_{\rm B}T$, and the rate $\nu$ depends on ATP concentration
   and internal timescales that determine the dwell-time statistics of the motor. 
  
   An individual trajectory of a motor starting 
   from a reference state $X(0)=0$ at time $t=0$ is denoted by 
   $X_{[0,t]} = \{X(s)\}_{s=0}^{t}$. It contains jumps $j=1,2,\dots$  from state $x_j^-$ to state $x_j^+$
   that occur at random times $t_j$. 
   The entropy production in units of $k_{\rm B}$ associated with this trajectory is
 $S(t)=\ln [\mathcal{P}(X_{[0,t]})/\mathcal{P}(\tilde{X}_{[0,t]})] = A X(t)$~\cite{neri2017statistics}. Here $\mathcal{P}$ is the path probability and $\tilde{X}_{[0,t]}= \{X(t-s)\}_{s=0}^{t}$ is the time reversed path. Thus,  the entropy production 
 $S(t)$ is a stochastic variable that undergoes 
 a  biased random walk of step size $A$ with trajectories $S_{[0,t]}=AX_{[0,t]}$.
  For $A$ positive, both the average velocity $v=\langle X(t)\rangle/t = (k_+ - k_-)=2\nu\sinh(A/2)$ and the average 
  rate of entropy production $\sigma = \langle S(t)\rangle/t = vA$ are positive. Here and in the following we denote by $\langle \,\cdot\, \rangle$ averages over many realizations of the process $X(t)$. 
  However, due to fluctuations, the stochastic variables $X(t)$ and  $S(t)$ can in principle take any value with
  finite probability and even become negative.

We  now derive exact expressions for the 
statistics of the   minimum $X_{\rm min}(t) = \text{min}_{\tau \in [0,t]} X(\tau)$  and the  maximum  $X_{\rm max}(t) = \text{max}_{\tau \in [0,t]} X(\tau)$ of the position of the motor with respect to its initial position, see Fig.~\ref{fig:1}b for  illustrations. We also discuss the global minimum and maximum of the
 stochastic entropy production 
$S(t)=AX(t)$ denoted by $S_{\rm min}(t)$ and  $S_{\rm max}(t)$, respectively. 
 We first discuss the statistics of the global extrema of the position $X_{\rm min}\equiv \lim_{t\to\infty}X_{\rm min}(t) $ and  
 $X_{\rm max}\equiv \lim_{t\to\infty}X_{\rm max}(t) $, and of the entropy production,  $S_{\rm min}$ and $S_{\rm max}$. The probability that the global minimum of the discrete position is $-x$, with $x\geq 0$, is $P(X_{\rm min} = -x) = P_{\rm abs}(-x) -  P_{\rm abs}(-x-1)$, 
where  $P_{\rm abs}(-x)=e^{-Ax}$~\cite{neri2017statistics} is the probability that $X(t)$ reaches an absorbing site in $-x$
  at  a finite time.  
Thus,  the global minimum follows a geometric distribution
 \begin{eqnarray}\label{eq:PSmin}
P(X_{\rm min}  = -x) &=&P(S_{\rm min}  = -Ax) \nonumber\\
&=&e^{-A x} (1-e^{-A})\quad,
 \end{eqnarray}
 for $x\geq 0$ and $P(X_{\rm min}  = x)=P(S_{\rm min}  = Ax)=0$ for $x>0$. 
 From Eq.~\eqref{eq:PSmin} we obtain the mean  global minimum of a 1D biased random walk and of its associated entropy production: 
  \begin{equation}\label{eq:Smin}
 \langle X_{\rm min} \rangle = \frac{-1}{e^A-1}\quad,\quad\langle S_{\rm min} \rangle = \frac{-A}{e^A-1}\quad.
 \end{equation}
 Therefore, the global minimum of the position diverges in the limit of small bias $A$ whereas the entropy production minimum is bounded for all $A\geq 0$ and  obeys the infimum law $ \langle S_{\rm min} \rangle \geq -1$~\cite{neri2017statistics}. This bound is saturated in the limit of small affinity, which corresponds to the diffusion limit~\cite{pigolotti2017generic}. Because $S(t)$ and $X(t)$ have positive drift, the average global maxima of entropy production and displacement are not defined. 
 However the difference $\lim_{t\to \infty}[\langle S_{\rm max}(t)\rangle-\langle S(t)\rangle] = A/(e^A-1)$ is finite 
 and obeys symmetry properties that we discuss below.

Finite-time extrema statistics of the 1D biased random walk may be obtained from the finite-time absorption probabilities 
\begin{eqnarray}\label{eq:Psmint}
P(X_{\min}(t)=-x)&=& P(S_{\min}(t)=-Ax) \\
&=& P_{\rm abs}(-x;t)-P_{\rm abs}(-x-1;t)\quad,\nonumber
\end{eqnarray}
where $P_{\rm abs}(-x;t)$ is the probability that $X$ reaches an absorbing site at $-x$ at any time smaller or equal than $ t$. The absorption probability $P_{\rm abs}(x;t)=\delta_{x,0}+\int_0^t   P_{\rm fpt}(\T;x)\d \T $, with $\delta_{i,j}$  Kronecker's delta and
\begin{equation}
 P_{\rm fpt}(\T;x)=e^{Ax/2}\frac{|x|}{\T}I_x(2\nu \T)e^{-2\nu\cosh(A/2)\T}	\quad,
 \label{fptd1}
 \end{equation}
is  the first-passage time probability  for the motor 
 to first reach an absorbing site in $x$, with $|x|\geq 1$, at time $\T\geq 0$~\cite{redner2001guide,roldan2016stochastic}, see Appendix~\ref{app:a}. Here 
 $I_x$ denotes the $x-$th order modified Bessel function of the first kind.  Note that $\int_0^{\infty} \text{d}\T   P_{\rm fpt}(\T;x) = P_{\rm abs}(x)\leq 1$. 
We identify in Eq.~(\ref{fptd1}) two timescales. 
The smaller timescale $\tau_1 =(k_++k_-)^{-1}= (2\nu \cosh(A/2))^{-1}$ is the average waiting time between two jumps, and $\tau_2=(2\sqrt{k_+k_-})^{-1}=(2\nu)^{-1}$ is inversely proportional to the geometric mean of the transition rates; their ratio $\tau_2/\tau_1=\cosh(A/2)\geq 1$ increases with the bias strength. Normalizing~\eqref{fptd1} by $P_{\rm abs}(x)$, we obtain  the mean $\langle \T \rangle = |x|A/\sigma$ and variance $\text{Var}[\T]=(\coth (A/2)/|x|)\langle \T \rangle^2$ of the first-passage time, in agreement with the first-passage time uncertainty relation $\text{Var}[\T]/\langle\T\rangle \geq 2/\sigma$~\cite{gingrich2017fundamental}.
Furthermore, the first-passage time probability density~\eqref{fptd1} obeys the following symmetry properties. First, the ratio 
\begin{equation}
P_{\rm fpt}(\T;x)/  P_{\rm fpt}(\T;-x)=e^{Ax}\quad,
\end{equation}  is independent on $\T$, as follows from the stopping-time fluctuation theorem~\cite{neri2017statistics,qian2006generalized,singh2019universal}. Second, the ‘‘conjugate" first-passage time probability $\tilde{P}_{\rm fpt}(\T;x)$, obtained exchanging $k_+$ by $k_-$ (i.e. $A$ by $-A$), obeys 
\begin{equation}
{P}_{\rm fpt}(\T;x)/\tilde{P}_{\rm fpt}(\T;x) = e^{Ax}\quad.
\end{equation}  These two properties imply $\tilde{P}_{\rm fpt}(\T;x) =P_{\rm fpt}(\T;-x)$, which has interesting consequences for random walks~\cite{krapivsky2018first} and for the extrema statistics of $S(t)$, see below.

\begin{figure}[ht!]
\includegraphics[width=8cm]{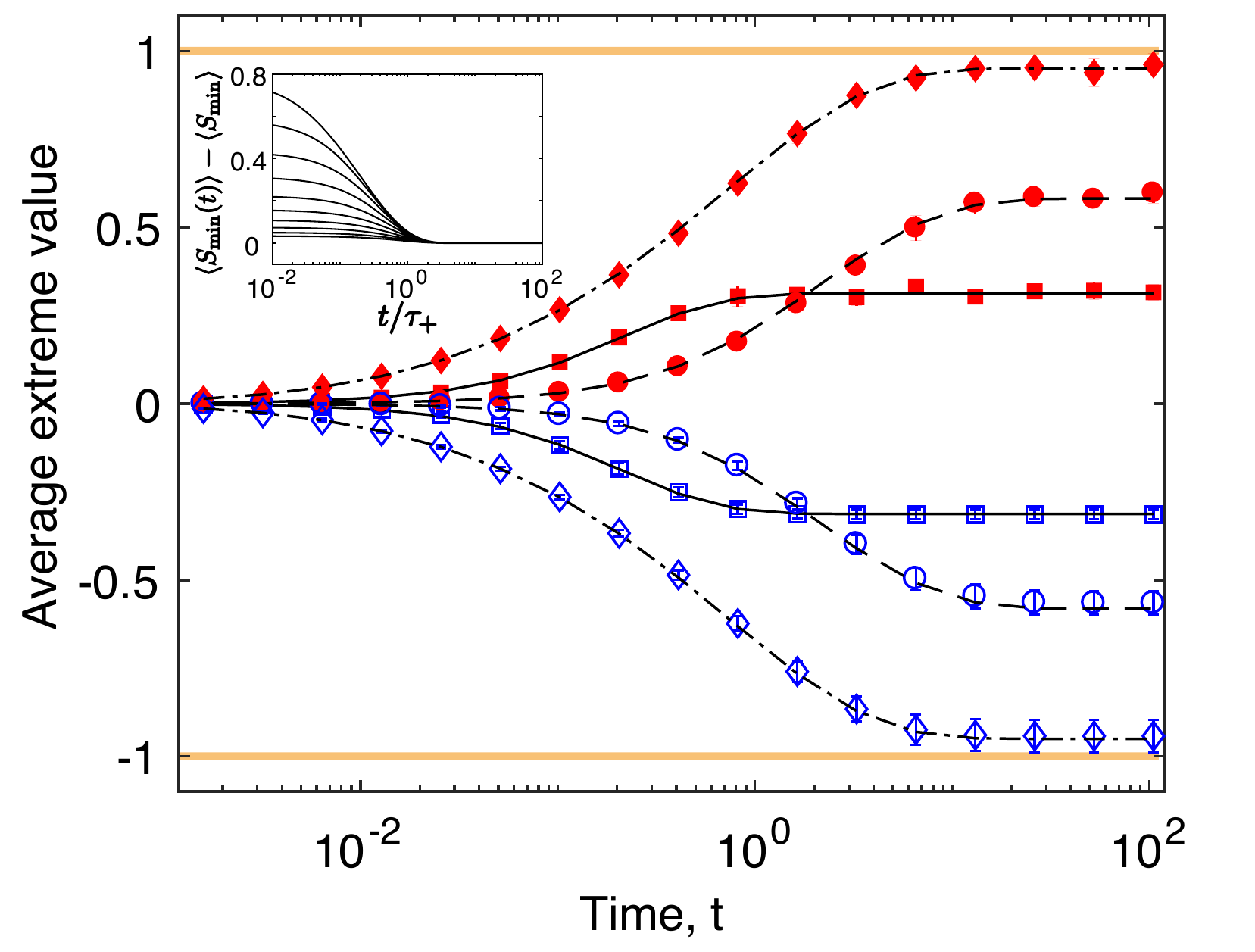}
\caption{Average minimum $\langle S_{\rm min}(t)\rangle$ (blue open symbols)  and average of the maximum minus the final value $\langle S_{\rm max}(t)\rangle-\langle S(t)\rangle$   (red filled symbols) of stochastic entropy production as a function of time of a 1D biased random walk. The symbols are averages over $10$ sets  of $10^3$ numerical simulations; the error bars are the standard deviation of the mean values obtained from these sets. The black lines are obtained from numerical integration of  Eq.~\eqref{eq:integral} using the trapezoidal method. 
Simulation parameters:  $A=1$, $\nu=0.5$ (squares);  $A=2$, $\nu=2$ (circles); $A=0.1$, $\nu=100$ (diamonds). The horizontal orange lines at $\pm 1$ correspond to the bound obtained using martingale theory~\cite{neri2017statistics}. Inset: $\langle S_{\rm min}(t)\rangle-\langle S_{\rm min}\rangle$ as a function of time rescaled by $\tau_\infty=[2\nu (\cosh(A/2)- 1)]^{-1}$. The different curves correspond to $\nu=1$ and $0.5 \leq A\leq 5$. }
\label{fig:2}
\end{figure}

\begin{figure}[ht!]
\includegraphics[width=7.5cm]{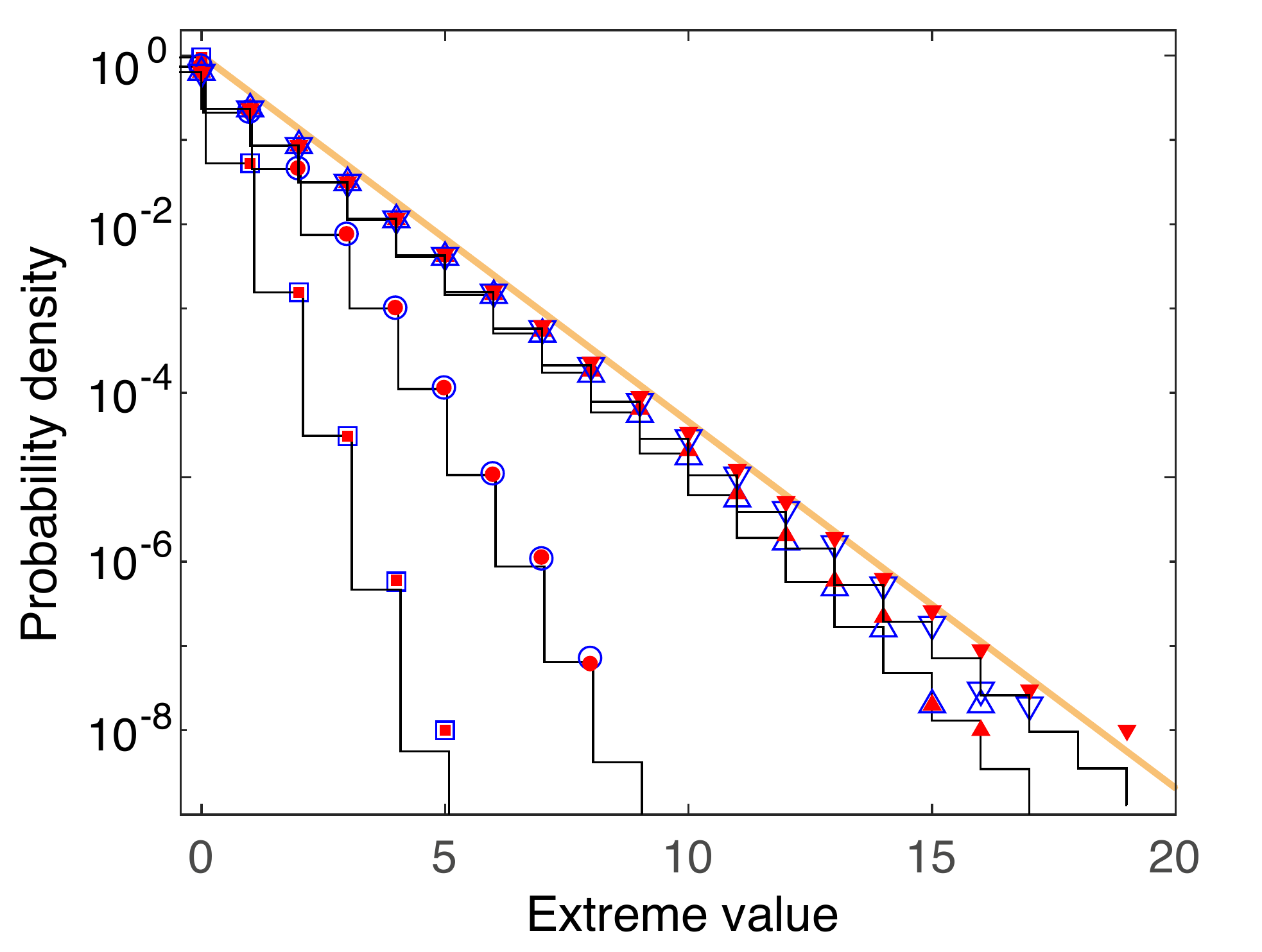}
\caption{Empirical probability density of $-S_{\min}(t)$ (blue open symbols) and $S_{\max}(t)-S(t)$ (red filled symbols) obtained from $10^8$ numerical simulations of a 1D biased random walk with parameters $A=1$ and $\nu=1$. Different symbols represent different integration times  $t=10^{-2}$ (squares), $t=10^{-1}$ (circles), $t=1$ (up triangles), $t=10$ (down triangles). The black lines are the theoretical distributions for different values of $t$ (from left to right) evaluated using Eq.~\eqref{eq:Psmint}. The orange line is an exponential distribution with mean value equal to one.\looseness-1 }
\label{fig:3}
\end{figure}

In order to derive exact finite-time extrema statistics, it is often  convenient to use generating functions and  Laplace transforms. The generating functions of the distributions of  finite-time entropy production extrema are defined as  $G_{\rm min/max}(z;t)=\sum_{x=-\infty}^{\infty}  z^{x} P\left(S_{\rm min/max}(t)=Ax\right)\Theta(\mp x)$, with $\Theta$ the Heaviside function.  Their Laplace transforms are given by $ \hat{G}_{\rm min/max}(z;s)=\int_0^{\infty}\text{d}t e^{-st}  G_{\rm min/max}(z;t) = s^{-1}[1-\hat{ P}_{\rm fpt}(s;-1)]/[1-z^{\mp 1}\hat{ P}_{\rm fpt}(s;-1)]$. Here, $\hat{ P}_{\rm fpt}(s;n)=e^{Ax/2}e^{-|x| \,\mathrm{cosh}^{-1} \left[s/2\nu+\cosh (A/2)\right]}-\delta_{x,0}$ is the Laplace transform of the first-passage-time probability density at site $x$.  
These expressions enable the computation of the Laplace transform of all the moments of the extrema from successive derivatives of the generating functions with respect to $\ln z$. In particular, the Laplace transform of the average minimum of entropy production reads $s\langle \hat{S}_{\min}(s)\rangle=-A/[\hat{ P}_{\rm fpt}(s;-1)^{-1}-1]$.
In the time domain, we may write this equality as $\langle S_{\min}(t) \rangle=-A\int_0^t \text{d}\T \sum_{x=1}^{\infty}P_{\rm fpt}(\T;-x)$~\cite{hartich2019extreme} which can be written as (see Appendix~\ref{app:b}):
\begin{equation}
\langle S_{\min}(t) \rangle=-\frac{A}{2\pi}\int_{-1}^1 \frac{1-e^{-2\nu t(\cosh(A/2)-y)}}{\left(\cosh(A/2)-y\right)^2}\sqrt{1-y^2}\,\d y\quad .
 \label{eq:integral}
\end{equation}
Numerical simulations of the 1D biased random walk are in  excellent agreement with Eq.~\eqref{eq:integral} (Fig.~\ref{fig:2}, blue symbols). Note that Eq.~\eqref{eq:integral}  can also be expressed in terms of the Kamp\'e de   F\'eriet function $\text{F}$ as $\langle S_{\min}(t)\rangle/A=
  -k_-t+{}^{2+0}F_{1+1}\left[\left.\genfrac{}{}{0pt}{}{ [2 \;1]\;,
  \;\varnothing\;, \; \varnothing}{ 3\;,\;2\;,\;2} \right\rvert -k_-t\;,\;
  -k_+t \right](k_-k_+t^2)/2$ (see Appendix~\ref{app:e}). Interestingly, our simulations  reveal (Fig.~\ref{fig:2}, red symbols) that the average maximum of entropy production minus the average entropy production at time $t$ equals to minus the right-hand side of Eq.~\eqref{eq:integral}:
\begin{equation}\label{eq:symmean}
\langle S_{\rm max} (t)\rangle -  \langle S(t)\rangle =- \langle S_{\rm min} (t) \rangle\quad.
\end{equation}
Equation~\eqref{eq:symmean} follows from the symmetry relation of the 1D biased random walk $\tilde{P}_{\rm fpt}(\T;x) =P_{\rm fpt}(\T;-x)$ and the relation 
 $\langle S_{\max}(t) \rangle - \langle S(t)\rangle=A\int_0^t \text{d}\mathsf{T} \sum_{x=1}^{\infty}\tilde{P}_{\rm fpt}(\mathsf{T};x) $. Moreover, this symmetry extends to the distribution of minima and maxima of entropy production
 \begin{equation}\label{eq:symdist}
P\left(S(0)-S_{\min}(t)=s\right)=P\left(S_{\max}(t)-S(t)=s\right) ,
 \end{equation}
 where, in this case, $S(0)=0$. 
 Figure~\ref{fig:3} shows empirical distributions of entropy-production minima and maxima obtained from numerical simulations, which fulfil  the symmetry relation~\eqref{eq:symdist}. 

\section{Random-matrix approach}
\label{sec3}
 We now  explore  a connection between entropy-production extrema  and random-matrix theory. More precisely, we relate the previously derived expressions for the average and distribution of extrema with eigenvalue distribution of specific random matrices. Equation~\eqref{eq:integral} can  also be written as (see Appendix~\ref{app:c})
 \begin{equation}\label{eq:MP}
 \langle S_{\rm min}(t)\rangle = \langle S_{\rm min}\rangle \left(1 - \int_{\tau_0}^{\tau_\infty} e^{-t/\tau}\rho(\tau/\bar{\tau})\frac{\d \tau}{\bar{\tau}} \right) \quad.
 \end{equation}
 Here $\tau_0 \equiv \left(\sqrt{k_+}+\sqrt{k_-}\right)^{-2}$ is the minimal relaxation time of the extreme value statistics and $\tau_\infty \equiv \left(\sqrt{k_+}-\sqrt{k_-}\right)^{-2}$ is the maximal extrema relaxation time. Here $\rho$ is the Mar{\v{c}}enko-Pastur distribution, where times are normalized by $\bar{\tau} \equiv k_+/(k_+-k_-)^2$. The Mar{\v{c}}enko-Pastur distribution is given by~\cite{marvcenko1967distribution}:
 \begin{equation}
\rho(\lambda)\equiv 
 \begin{cases} \dfrac{1}{2\pi \delta}\dfrac{\sqrt{(\lambda_+-\lambda)(\lambda-\lambda_-)}}{\lambda} &\;\mbox{if}\;\; \lambda \in [\lambda_-,\lambda_+]\quad \\
0 & \;\mbox{if}\;\;\lambda \notin [\lambda_-,\lambda_+],\quad \end{cases} \label{eq:MPdefMT}
  \end{equation}
with $\int_0^\infty \rho(\lambda)\d \lambda =1$, where $\delta = k_-/k_+ = e^{-A}$ and $\lambda_\pm=\left(1\pm\sqrt{\delta}\right)^2 = (\sqrt{k_+}\pm\sqrt{k_-})^2/k_+$.
Interestingly,  $\rho(\lambda)$ is the   distribution of eigenvalues in the large size limit of Hermitian matrices drawn from the ensemble of the Wishart-Laguerre random matrices~\cite{nica2006lectures,fridman2012measuring,livan2018introduction}, whose structure is explained below. The average~\eqref{eq:MP} follows from the generating function of the distribution of the minimum (see Appendix~\ref{app:c})
 \begin{eqnarray}\label{eq:GFext}
\hspace{-0.5cm}G_{\min}(z;t)=1+\frac{1-z}{e^A-1}\int_{\tau_0}^{\tau_\infty} \frac{1-e^{-t/\tau}}{1+f(z)\tau}\rho(\tau/\bar{\tau})\frac{\d \tau}{\bar{\tau}}\;, \end{eqnarray}
where $f(z)=k_+(z-1)+k_-(z^{-1}-1)$. 

Equations~(\ref{eq:MP}-\ref{eq:GFext}) imply that the time at which the distribution of the extrema relax to their long time limit is given by the largest timescale of the relaxation spectrum $\tau_\infty =\bar{\tau}\lambda_+$. We demonstrate this result in an example shown in the inset of Fig.~\ref{fig:2} which shows this for the case of $\langle S_{\rm min}(t)\rangle$.   Furthermore, it implies that the following trace formula holds: 
\begin{equation}\label{SminRMT}
\langle S_{\rm min}(t)\rangle =  \langle S_{\rm min} \rangle  \left( 1 - \lim_{m\to \infty} \tfrac{1}{m}\Tr
e^{-  \mathbf{M}^{-1} t/\bar{\tau}} \right)\quad,
\end{equation}
%in the limit of large matrix size $m$, 
where $\mathbf{M}$ is a $m\times m$ random matrix. 
Eq. (\ref{SminRMT}) holds for $\mathbf{M}$ drawn from one of two random matrix ensembles: 
(i) Wishart matrices $\mathbf{M}=n^{-1}\mathbf{R} \mathbf{R}^T$, where $\mathbf{R}$ is a $m\times n$ rectangular random matrix, $\mathbf{R}^T$ its transpose 
where $n=\lceil e^A m\rceil$. Its entries $R_{ij}$ are  independent and identically distributed Gaussian random numbers with 
zero mean and unit variance;  (ii) Laguerre matrices $\mathbf{M}=n^{-1}\mathbf{R} \mathbf{R}^T$,  where $\mathbf{R}$ is  a $m\times m$ square matrix with 
$R_{ij}$  independent but not identically distributed random variables, drawn from specific $\chi$ distributions~\cite{dumitriu2002matrix}, see Appendix~\ref{app:d}. 
%scaled by its degree of freedom $n=e^Am$. Note that for a finite matrix size $m$, we obtain an estimator for $\langle S_{\rm min}(t)\rangle$. Both the Whishart and the Laguerre random matrix can be generated as the product of a random matrix and its transpose: $n^{-1}\mathbf{R} \mathbf{R}^T$. In the Whishart case, $\mathbf{R}$ is a $m\times n$ rectangular random matrix~\cite{dumitriu2002matrix} (with $n$ rounded to the nearest integer), whose entries $R_{ij}$ are independent and identically distributed random variables drawn from a Gaussian distribution of zero mean and unit variance. In the Laguerre case, $\mathbf{R}$ is a $m\times m$ tridiagonal square matrix with $R_{ij}$ independent but not identically distributed random variables, each given by the square root of a random number drawn from a chi square distribution, see Appendix~\ref{app:d}. } 
%{\color{gray}Furthermore, it suggests that one can approximate the average minimum as 

Eq. (\ref{SminRMT}) can be approximated numerically using random matrices with finite but sufficiently large $m$. In practice,
we use the following estimate:
\begin{equation}%\label{SminRMT}
\langle S_{\rm min}(t)\rangle \simeq  \langle S_{\rm min}\rangle \left( 1 - \frac{1}{m}\sum_{i=1}^m 
e^{- t/(\bar{\tau} \lambda_i)} \right)\quad,
\end{equation}
where $\lambda_i$ is the $i$-th eigenvalue of $\mathbf{M}$ drawn from either the Wishart or Laguerre ensembles.

\begin{figure}
\includegraphics[width=7.5cm]{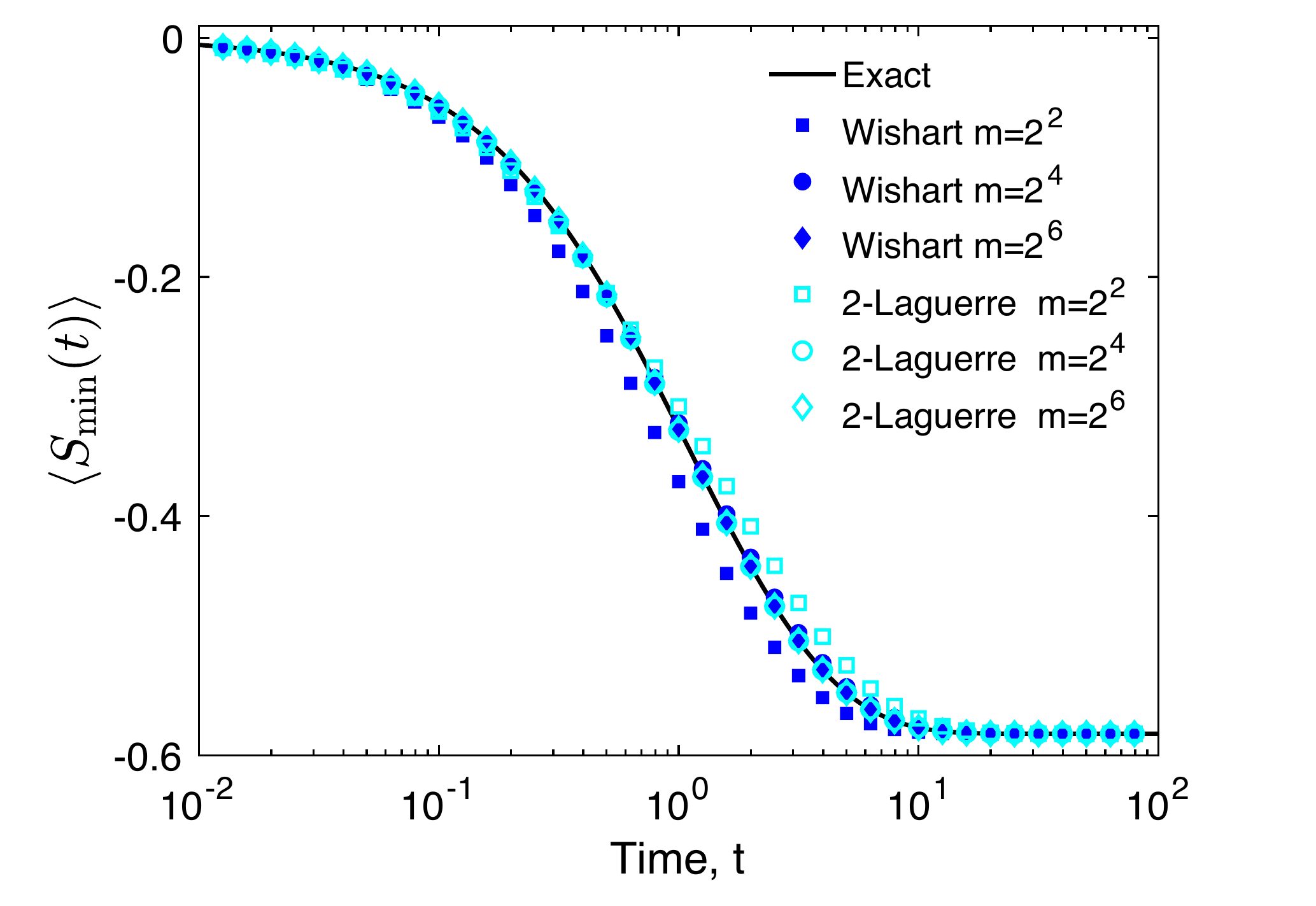}
\caption{Finite-time average minimum of entropy production associated with 1D biased random walks as a function of time: exact result   (Eq.~\eqref{eq:MP}, black line) and estimates obtained from the spectrum of \textit{a single} $m\times m$ random matrix drawn from the Wishart (blue filled symbols) and the $\beta$-Laguerre ensembles ($\beta=2$, open cyan symbols). The random-matrix estimates are obtained evaluating the right-hand side in Eq.~\eqref{SminRMT}, i.e. $\langle S_{\rm min}\rangle [ 1 - m^{-1}\sum_{i=1}^m \!e^{-t/(\bar{\tau}\lambda_i)} ]$ with $\langle S_{\rm min}\rangle=-A/(e^A -1)$, $\bar{\tau} \equiv k_+/(k_+-k_-)^2=e^{A/2}/(4\nu\sinh^2(A/2))$, and $\lambda_i$ the $i$-th eigenvalue of the corresponding Wishart/Laguerre random matrix. Values of the parameters: $A=\nu=1$.}
\label{fig:4}
\end{figure}

 In Fig.~\ref{fig:4} we illustrate the validity of our random-matrix approach. More precisely, we compare the theoretical value of the average minimum of entropy production with the estimate given by the right-hand side of Eq.~\eqref{SminRMT}. For this purpose, we compute numerically the eigenvalues of a single random matrix of the Wishart ensemble and of a matrix drawn from the $\beta-$Laguerre ensemble with parameter $\beta=2$.  Notably, using a single $64\times64$ random matrix from the Wishart or Laguerre ensembles, we obtain an estimate of the average entropy production minimum that  differs with respect to the exact value only by about $2\%$ at all  times.\looseness-1

\begin{figure}
\includegraphics[width=7.5cm]{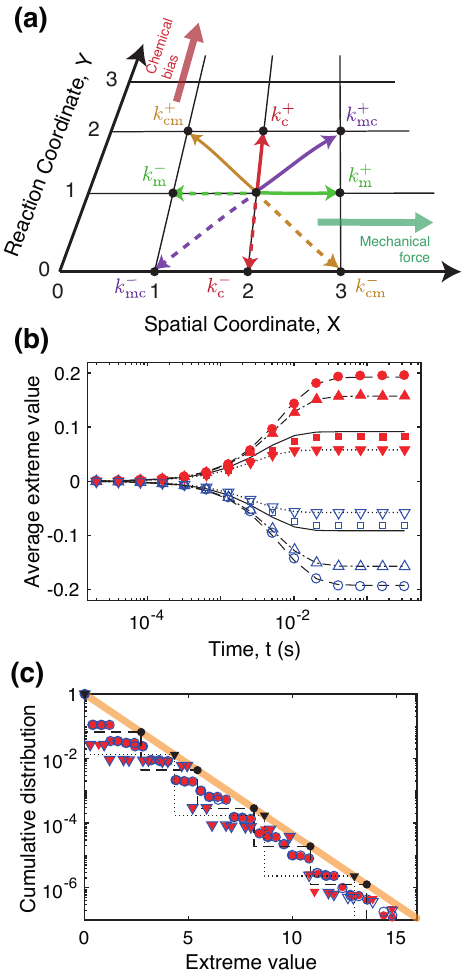}
\caption{Entropy production extrema for a two-dimensional stochastic model of a molecular motor.  (a) Sketch of the model with states given by the vertices of the 2D grid and the possible transitions from state (2,1) marked with  arrows. (b) Finite-time average minimum of entropy production ($\langle S_{\rm min}(t)\rangle$, blue open symbols) and average maximum of the entropy production minus its final value ($\langle S_{\rm max}(t)-S(t) \rangle$, red filled symbols) for values of the external force $f_{\rm ext}=-2.5$pN (squares, solid line), -1.5pN (circles, dashed line), -0.5pN (up triangles, dash-dot line), 0.5pN (down triangles, dotted line). The lines are estimates obtained using  Eq.~\eqref{eq:integral} with effective parameters $A_{\rm eff}$ and $\nu_{\rm eff}$ given by Eqs.~(\ref{eq:nueff}-\ref{eq:Aeff}), see text for further details.  (c) Cumulative distribution of  $-S_{\rm min}(t)$  (blue open symbols) and of $S_{\rm max}(t)-S(t)$ (red filled symbols) for  $f_{\rm ext}=-1.5$pN (circles) and 0.5pN (down triangles), and   $t=50\,\rm ms$. The black symbols are estimates given by   Eq.~\eqref{eq:PSmin} with effective parameters $A_{\rm eff}$, and the orange line is an exponential distribution with mean one. Values of the simulation parameters: $\kb T=4.28$pN$\cdot$nm, $\ell = 8$nm, $\Delta \mu = 4\kb T$, $\nu_\textrm{m}=10$Hz, $\nu_\textrm{c}=5$Hz, $\nu_\textrm{mc}=25$Hz, and $\nu_\textrm{cm}=1$Hz. The numerical data were obtained from $10^8$ simulations done using Gillespie's algorithm.} %Values of parameters: $\kb T=4.28$pN$\cdot$nm, $\ell = 8$nm, $\Delta \mu = 4\kb T$, $\nu_\textrm{m}=10$Hz, $\nu_\textrm{c}=5$Hz, $\nu_\textrm{mc}=25$Hz and $\nu_\textrm{cm}=1$Hz. The distributions in (c) and the mean in (b) were obtained from $10^8$ simulations done using Gillespie algorithm.}}
\label{fig:5}
\end{figure}

\section{Extrema statistics of molecular motors}% with broken detailed balance
\label{sec4}
 We now investigate whether similar results also hold for more complex stochastic models of molecular motors. We consider a biochemical process where a molecular motor's  fluctuating motion is described by a continuous-time Markov jump process on a  potential energy surface in two dimensions $x$ and $y$ (Fig.~\ref{fig:5}a). Here $x$ denotes the spatial displacement of the motor along a discrete track of period $\ell$, and  $y$ is  a chemical reaction coordinate denoting the net number of fuel molecules spent by the motor. 
 
The motion of the motor is biased along the track by a mechanical force $f_{\rm ext}$ applied to the motor. In addition, the motor hydrolyzes ATP with  chemical potential difference $\Delta \mu$. We consider that both $f_{\rm ext}$ and $\Delta \mu$ to be independent of the state of the motor, which corresponds to the limits where the external force and the concentration of fuel molecules are stationary.  States $(x,y)$ of the motor are in local equilibrium at temperature $T=\beta^{-1}$. The dynamics of the motor is as follows.
 From a given state  the motor can perform, at a random time, a jump to eight adjacent states corresponding to the following four transitions and their reversals: (i) sliding along the track by a distance 
 $\ell$ ($-\ell$) without consuming fuel but generating work in (against) the direction of the force,  at a rate $k_{\rm m}^{+}$ ($k_{\rm m}^{-}$), with $k_{\rm m}^{-} = k_{\rm m}^{+} e^{-\beta f_{\rm ext} \ell}$; (ii) consumption of one ATP (ADP) molecule without generating work,   at a rate $k_{\rm c}^{+}$ ($k_{\rm c}^{-}$), with $k_{\rm c}^{-} = k_{\rm c}^{+} e^{-\beta\Delta \mu}$; (iii) work generation in (against) the external force using ATP (ADP) at a rate   $k_{\rm mc}^{+}$ ($k_{\rm mc}^{-}$), with  $k_{\rm mc}^{-} = k_{\rm mc}^{+} e^{-\beta(f_{\rm ext}\ell+\Delta \mu)}$
and (iv) work generation against (in) the external force using ATP (ADP) at a rate   $k_{\rm cm}^{+}$ ($k_{\rm cm}^{-}$), with  $k_{\rm cm}^{-} = k_{\rm cm}^{+} e^{\beta(f_{\rm ext}\ell-\Delta \mu)}$. We use transition rates of the form $k_{\alpha}^{\pm} = \nu_{\alpha} e^{\pm A_\alpha/2}$, where $\nu_\alpha$ give different weights to each transition type.
%with $\nu_{\alpha}$. 
%a rate that we assume, for simplicity, to be independent on 
%the transition type $\alpha$.\looseness-1
 
 A single trajectory of the motor is a 2D random walk containing snapshots $(X(t),Y(t))$ of the state of the motor at time $t$. Here $X(t)$ is the spatial coordinate of the motor (with respect to its initial position) and $Y(t)$ is the reaction coordinate representing the net number of ATP molecules consumed up to time $t$. Note that when $Y(t)$ is negative, the motor has consumed more ADP than ATP molecules.  
 The entropy production 
 associated with  a single trajectory of the molecular motor is $S(t)=A_{\rm m}X(t) + A_{\rm c}Y(t)$, 
 where $A_{\rm m}=\beta f_{\rm ext}\ell$, 
 $A_{\rm c}=\beta \Delta \mu$ are the mechanical and chemical affinities.
 Thus $S(t)$ is a random walk with four different step lengths $A_{\rm m}$, $A_{\rm c}$, $A_{\rm mc}\equiv A_{\rm m}+A_{\rm c}$ and $A_{\rm cm}\equiv -A_{\rm m}+A_{\rm c}$ corresponding to the jumps along the $X$, $Y$ and the diagonal directions, respectively.
%\textcolor{orange}{ Transitions  from states $(x,y)$ to $(x\pm 1,y\pm1)$  induce broken detailed balance as they can occur due to simultaneous or sequential steps of  work generation and chemical energy expenditure.  \textit{This remark was not clear for referee 3.}}
 
\vspace{1cm}
 We perform numerical simulations of this 2D stochastic model of the molecular motor using Gillespie's algorithm, and evaluate the entropy flow associated with different trajectories of the motor.  Obtaining exact extreme-value statistics in this model is challenging. However the following simple approximation provides good estimates.  
 The  finite-time average   (Fig.~\ref{fig:5}b) and the distribution  (Fig.~\ref{fig:5}c) of the  entropy production extrema 
 obtained from simulations can be approximated by Eq.~\eqref{eq:integral} and Eq.~\eqref{eq:PSmin} replacing $A$ and $\nu$ by the effective parameters
 \begin{eqnarray}
 \nu_{\rm eff}&=&\sum_\alpha \nu_\alpha\quad, \label{eq:nueff}\\
 A_{\rm eff}&=&2 \mathrm{cosh}^{-1}\left(\sum_\alpha \frac{\nu_\alpha}{ \nu_{\rm eff}}\cosh \frac{A_\alpha}{2}\right)\quad,\label{eq:Aeff}
 \end{eqnarray}
 where the index $\alpha$ runs over the four types of transitions $\alpha=\text{m}$, $\text{c}$, $\text{mc}$
 and $\text{cm}$.
 This approximation based on effective parameters follows from considering effective 1D models with jumping rate $k^{+}_{\rm eff}+k^{-}_{\rm eff}=\sum_{\alpha}k^{+}_\alpha+k^{-}_\alpha$. Moreover, the symmetry between the time-dependent distribution of the extrema~\eqref{eq:symdist} is yet satisfied with high accuracy in our numerical simulations, even though these distributions can have very irregular shapes depending on the ratios of the affinities (Fig.~\ref{fig:5}c). 
 
  \begin{figure}
\includegraphics[width=7.5cm]{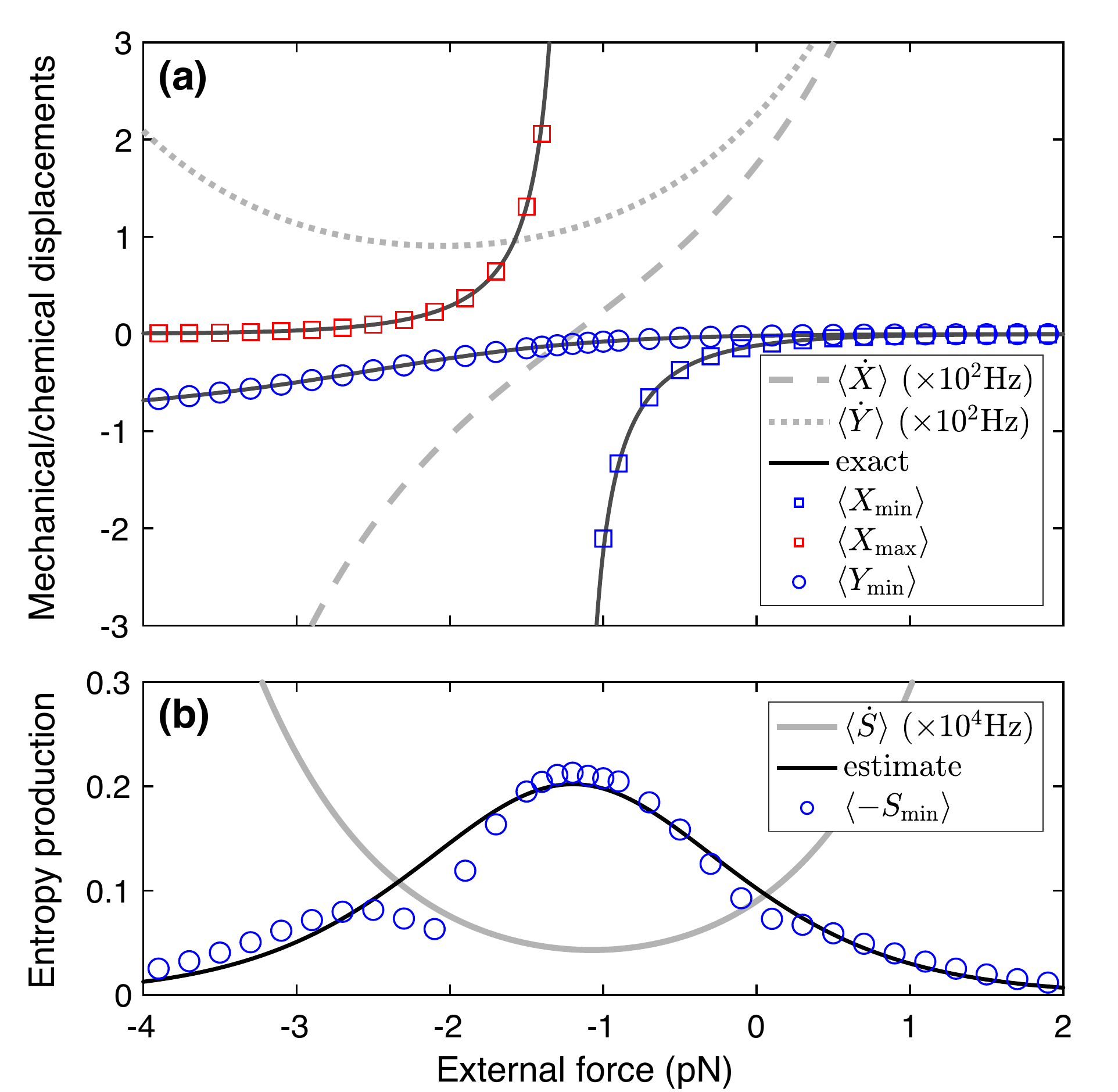}
\caption{Numerical results of mechanical and chemical currents and of entropy production extrema as a function of the force for the two-dimensional stochastic model of a molecular motor of Fig.~\ref{fig:5}.  (a) Average long time minima and maxima of mechanical steps $\langle X_{\min/\max}\rangle$ (blue and red squares respectively) and average minima of chemical steps at long time $\langle Y_{\min}\rangle$ (blue circle) are shown as a function of the external force. The black lines are analytical results obtained using Eq.~\eqref{eq:Smin} with effective affinities $A_{\rm x/y}=\ln (k_{\rm x/y}^+/k_{\rm x/y}^-)$. For comparison, we also show the mechanical stepping rates $k_{\rm x}^+-k_{\rm x}^-$ (dashed grey line) and the chemical stepping rate $k_{\rm y}^+-k_{\rm y}^-$ (dotted grey line). $k_{\rm x/y}^{\pm}$ are defined in Eq.~\eqref{xyrates}. (b) Average long time minima of total entropy production (blue circles) and rate of total entropy production (grey line) as a function of the external force. The black line is an estimation obtained using Eq.~\eqref{eq:Smin} with the effective parameters $A_{\rm eff}$ and $\nu_{\rm eff}$ given by Eqs.~(\ref{eq:nueff}-\ref{eq:Aeff}). The values of the fixed parameters are the same as in Fig.~\ref{fig:5}. The averages at long time in (a) and (b) were obtained at $t=0.5$s from $10^6$ simulations done using Gillespie algorithm.} %, where the extrema has converged to the long time value as can be observed in Fig.~\ref{fig:5} 
\label{fig:6}
\end{figure}
   
 %for the total entropy production 
The average extrema of the mechanical and chemical currents are shown in Fig.~\ref{fig:6} as a function of the external force. The behavior of these currents can be understood from a mapping to 1D biased random walks $X(t)$ and $Y(t)$ with effective forward and backward hopping rates given, respectively, by
\begin{align}\label{xyrates}
k_{\rm x}^{\pm}&=k_{\rm m}^{\pm}+k_{\rm mc}^{\pm}+k_{\rm cm}^{\mp} , \quad  k_{\rm y}^{\pm}=k_{\rm c}^{\pm}+k_{\rm mc}^{\pm}+k_{\rm cm}^{\pm} \quad.
\end{align}
This yielding effective affinities $A_{\rm x/y}$ and rates $\nu_{\rm x/y}$ defined as in Eq.~\eqref{eq:2}. This allows us to calculate the extreme value statistics of the numbers of step $X(t)$ and $Y(t)$, as illustrated in Fig.~\ref{fig:6} (a) as solid lines.
Note that $A_{\rm x}=\beta (f_{\rm ext}- f_{\rm stall})\ell$ is related to the mechanical affinity $A_{\rm m}=\beta f_{\rm ext}\ell$ and the stall force
\begin{align}
 f_{\rm stall}=\frac{1}{\beta \ell} \ln \left(\frac{1+k_{\rm mc}^{-}/k_{\rm m}^{-}+k_{\rm cm}^{+}/k_{\rm m}^{-}}{1+k_{\rm mc}^{+}/k_{\rm m}^{+}+k_{\rm cm}^{-}/k_{\rm m}^{+}}\right) \quad .
\end{align}%by an constant independent of the external force $f_{\rm ext}$
In the examples shown in Figs.~\ref{fig:5},\ref{fig:6}, $f_{\rm stall}\simeq-1.2$pN in the simulations. $A_{\rm y}$ and the chemical affinity $A_{\rm c}$ obey a similar relationship.
We observe in Fig.~\ref{fig:6} that the largest average extreme values of entropy production occur when the external force applied to the motor is near the stall force. Interestingly, the minima and maxima of the displacements, $X_{\min}$ and  $ X_{\max}$, show diverging averages when the stall force is approached from above or below respectively. Such behavior could be observable in experiments on the statistics of the stepping of molecular motors near stall forces. Note that in Fig.~\ref{fig:6} the long time limit of the average extrema is already reached at $0.1$s, see Fig.~\ref{fig:5}.
%$\langle X_{\min} \rangle

\section{Illustration of extreme value statistics using data from sports}
\label{app:sport}

Our analysis of extreme values in Sec.~\ref{sec2} applies to any stochastic  
process described by a biased random walk, e.g.  scores during a sports match.  Let us consider a game where two teams (Team 1 and Team 2) play against each other.
% scoring points at a certain average rate. This condition suits several sports with finite duration and time-homogeneous scoring rates, such as: soccer, basketball, hockey, handball, etc.      We consider here a minimal model for a  sports  match that does not take into account correlation non-stationary effects~\cite{Karlis03}. The results presented here are however sufficient to illustrate basic  principles to design  an exciting game in different sport disciplines.
% The results presented here are however sufficient to illustrate basic design principles for an exciting game.
 We denote the number of points scored by each team up to time $t$  by
 %by two Poisson processes 
 $N_1(t)$ and $N_2(t)$, respectively. We can define the average rate of scores for Team 1 
and Team 2 as  $k_+$ and $k_-$, where we consider without loss of generality $k_+>k_-$.
Here $k_{\pm}$ are the expected number of points scored per unit of time (e.g. goals per minute). 
 Their difference  $X(t)=N_1(t)-N_2(t)$ is the net score in favor of Team 1, i.e. it is positive when Team 1 is winning, zero when the match is tied, and negative when Team 2 is winning. 
 When applying our minimal model we assume for simplicity that the score of each team follows Poissonian statistics, i.e. we do not take into account correlations and non-stationary effects~\cite{karlis2003analysis}. 
 
We  now discuss our results in the context of this simplified model of sports. 
At finite time $t$, the stronger team is expected to lead by 
$\langle X(t) \rangle=(k_+-k_-)t$, with variance $\text{Var}[X(t)]=(k_++k_-)t$. %However, surprises are what we adore in such games. 
The following questions then appear naturally: 
``By how many points do we expect the weaker team to lead over the stronger one during the game?'', ``How long should a match last in order to see an exciting game with comebacks of the weaker team?". We provide insights on these questions in terms of $X_{\min}(t)$ i.e.  the  maximum net score against the stronger team up to time~$t$. %At long times compared to the duration of a match,  the stronger team will always lead, and the average  (Eq.~\eqref{eq:Smin}): $\langle X_{\min} \rangle=-k_-/(k_+-k_-)$. 
Knowledge on the characteristic time of the score's extrema statistics is useful for the design of an exciting game that takes our breath away as long as possible:   the longer the game, the more certain the stronger team is to win but the more predictable is the game, whereas  the shorter the game, the more random  it is.  Our theory reveals a continuous spectrum of the characteristic relaxation times of the score extrema, following the Mar{\v{c}}enko-Pastur distribution~(\ref{eq:MP}-\ref{eq:MPdefMT}). An exciting game with comebacks  should last at least $\tau_0=\left(\sqrt{k_+}+\sqrt{k_-}\right)^{-2}$. This is because $\tau_0$ is the timescale for the first action in the
game. It should not  be longer than $\tau_\infty=\left(\sqrt{k_+}-\sqrt{k_-}\right)^{-2}$ because after that time  the weaker team is not expected to lead anymore. Indeed, $\tau_\infty$ corresponds to the time after which the weaker time falls behind for the rest of the match. Interestingly the smallest timescale $\tau_0$ is shorter than the expected time for the first point  $\tau_1=(k_++k_-)^{-1}$, which defines the maximum of the Mar{\v{c}}enko-Pastur density function. 
% From this it follows that the game should last at least ---the typical time for the first point--- but no longer than  $\tau_\infty=\left(\sqrt{k_+}-\sqrt{k_-}\right)^{-2}$ ---the time after which we do not expect the weaker team to lead anymore.   For completeness, the short timescale $\tau_0=\left(\sqrt{k_+}+\sqrt{k_-}\right)^{-2}$ can be understood as the characteristic time for an action, which could lead to a first goal or comeback of the weaker team. 
%Note that these arguments not only apply to the extreme value $-X_{\min}(t)$ but also to $X_{\max}(t)-X(t)$, which corresponds to the maximal lead of the stronger team as compared to the final score, see .}

%The symmetry of this model further implies that these quantities also applies to the difference between 
%the maximum leading of the stronger team and its actual leading.

\begin{figure}[h!] %$\rho(\tau/\bar{\tau})/\bar{\tau}$ is a function  %(in min${}^{-1}$, normalized when integrated with respect to the linear x-axis $\int\d\tau$) 
\includegraphics[width=8cm]{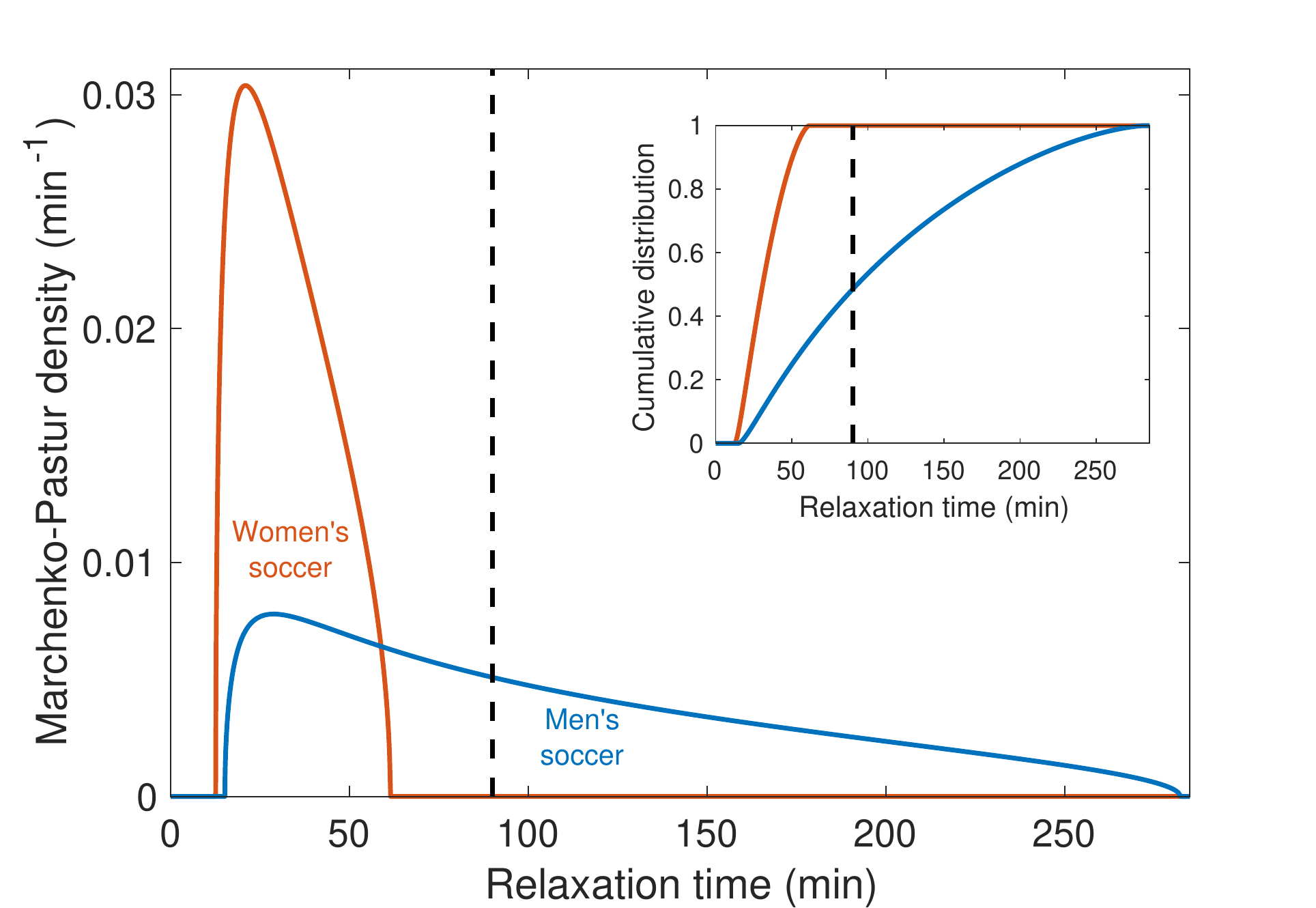}%M_vs_W_soccer.pdf}
\caption{Mar{\v{c}}enko-Pastur density $\rho(\tau/\bar{\tau})/\bar{\tau}$ as a function of the relaxation time $\tau$ for the extreme-value statistics corresponding to a 1D biased random walk, where $\bar{\tau}$ is a characteristic timescale. The distributions are estimated for scores of women's soccer (red line) and men's soccer (blue line), see parameters in Table~\ref{tab:1}. The inset shows the corresponding cumulative distributions. The vertical dashed line at 90min corresponds to the duration of a soccer match. The characteristic times are estimated in Table~\ref{tab:1}. The dashed line corresponds to the duration of the match.}\label{fig:7}
%\textcolor{blue}{Women soccer: peak at 22', cutoff at 68'. Men soccer: peak at  28', cutoff at 243'.   Inset: the median of men soccer is exactly at $90'$.}}
\includegraphics[width=8cm]{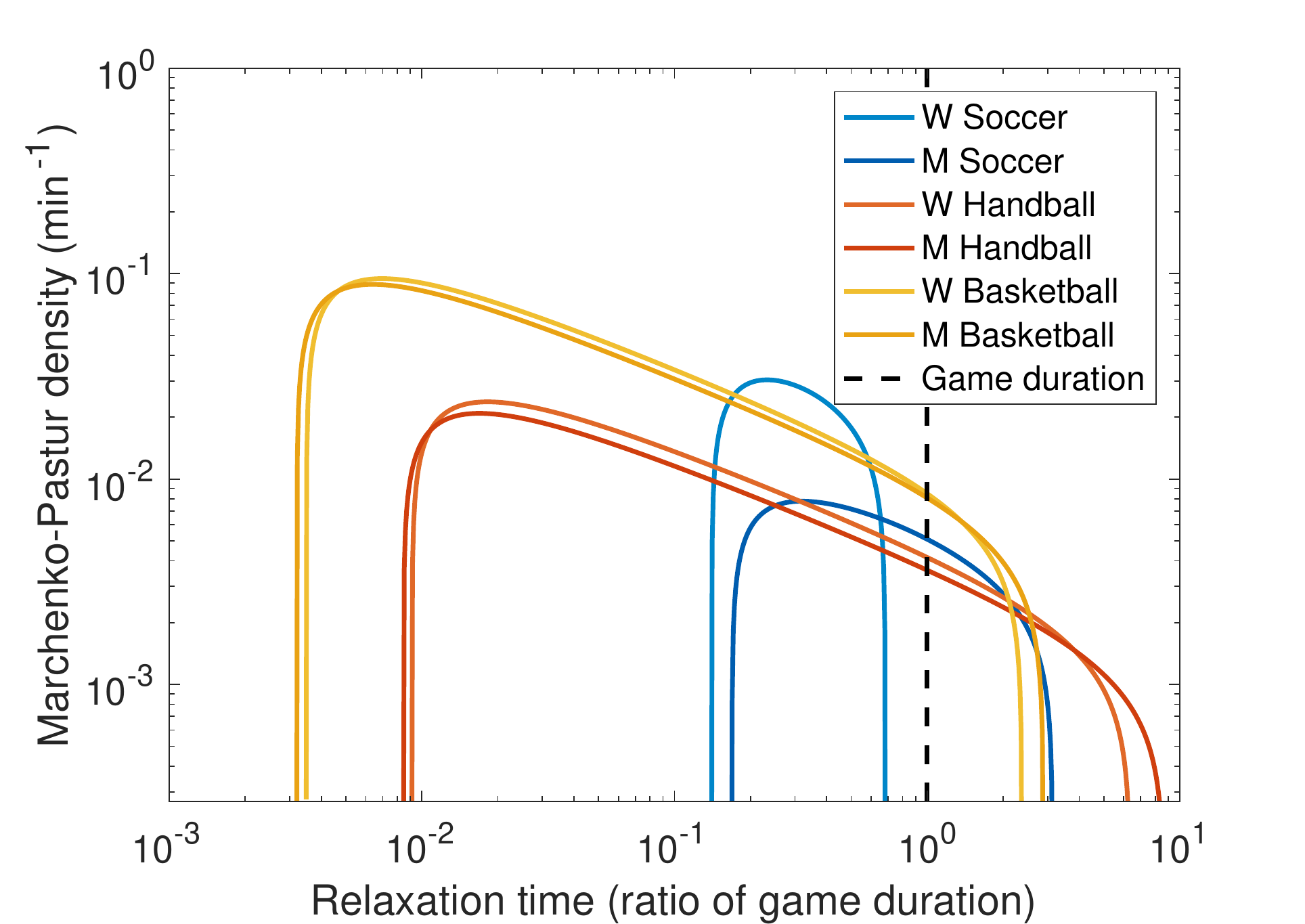}%M_vs_W_soccer.pdf}
\caption{Mar{\v{c}}enko-Pastur density $\rho(\tau/\bar{\tau})/\bar{\tau}$ as a function of the normalized relaxation time $\tau/\tau_{\rm match}$ for the extreme-value statistics corresponding to a 1D biased random walk, estimated for different sports.
Here $\tau_{\rm match}$ is the game duration of the respective sports. Parameters are: handball - women (men) $A_{\rm h} = 0.13\pm 0.04$  ($0.15\pm 0.04$), $\nu_{\rm h}=0.46\pm0.01$goals/min ($0.49\pm0.01$goals/min); basketball - women (men) $A_{\rm b}=0.13\pm0.02$  ($0.15\pm0.02$), $\nu_{\rm b}=1.79\pm0.02$points/min ($1.95\pm0.02$points/min); soccer as in Table~\ref{tab:1}. The dashed line corresponds to the duration of the match.}\label{fig:8}
\end{figure}

We illustrate these results using data from soccer, corresponding to the 2018-2019 season of Men's and Women's UEFA Champions League, see Table~\ref{tab:1}.  
We collect data of the total number of goals scored (GF) and goals conceded (GC) by all teams that qualified 1st at the end of the group stage.  We estimate the parameters  $k_+$ and $k_-$ for a typical match between the group leader against an average  team, see Appendix. 
\begin{table}[]
\begin{tabular}{l|c|c}
Soccer teams & Women & Men   \\
  \hline
  Goals scored GF & 113 & 108  \\  
  Goals conceded GC & 16 & 42 \\
  Number of matches MP & 30 & 48 \\
  \hline
$k_+$ (goals/min) & $0.042\pm0.006$ & $0.025\pm0.010$   \\
$k_-$ (goals/min) & $0.006\pm0.002$ & $0.009\pm0.002$   \\
  \hline
$A_{\rm s}=\ln(k_+/k_-)$ & $1.95\pm0.27$ & $0.94\pm0.18$  \\
$\nu_{\rm s}=\sqrt{k_+k_-}$ (goals/min)&$ 0.016\pm0.002$ & $0.016\pm0.001$   \\ 
  \hline
% $-\langle X_{\min}\rangle=k_-/(k_+-k_-)$ &  $0.16\pm0.05$ & $0.6\pm0.2$  \\
%  \hline
$\tau_0=\left(\sqrt{k_+}+\sqrt{k_-}\right)^{-2}$ (min) & $13\pm 1\;\;$  & $15\pm 1\;\;$  \\ 
$\tau_1=(k_++k_-)^{-1}$ 	(min) & $21\pm 2\;\;$ & $29\pm 2\;\;$  \\ 
$\bar{\tau}=k_+(k_+-k_-)^{-2}$ 	(min) & $32\pm 5\;\;$ & $110\pm 30\;\;$  \\ 
$\tau_\infty=\left(\sqrt{k_+}-\sqrt{k_-}\right)^{-2}$  (min) & $61\pm13$ & $280\pm100$ 
\end{tabular}
\caption{Number of goals scored and conceded by the teams qualified 1st at the end of the group stage in Women's (10 teams out of 40) and Men's (8 teams out of 32) UEFA Champions League in the 2018-2019 season, and their total number of matches. Estimation of the 1D model parameters from data of the season 2018-2019 of the soccer champions leagues (women and men). The scoring rates  are given by the goals scored and goals conceded divided by the number of matches and the duration of a match ($\tau_{\rm match}=90$min), i.e. $k_+=\text{GF}/(\tau_{\rm match}\text{MP})$ and $k_-=\text{GC}/(\tau_{\rm match}\text{MP})$. The rest of the parameters are obtained from our expressions corresponding to the 1D biased random walk: the scoring affinity $A_{\rm s}$ and rate $\nu_{\rm s}$ [Eq.~\eqref{eq:2}], and the extreme-value timescales $\tau_0$, $\tau_1$, $\bar{\tau}$ and $\tau_\infty$, see Sec.~\ref{sec3}. The uncertainties are obtained by propagating the standard error of the mean Poissonian numbers of goals per match.}
%The data consists in the cumulated number of goals for / against the best team of each group (during the group phase of the competition) and the corresponding number of match played .}}
\label{tab:1}
\end{table}
We find that the effective scoring affinity $A_{\rm s}=\ln (k_{+}/k_-)$, which measures the strength difference between group leaders and average soccer teams, is twice larger for Women's than for Men's soccer. This striking difference is not likely to be a fundamental difference in the way soccer is played for each gender, as suggested by their equal scoring rate $\nu_{\rm s}=0.016$goals/min. It rather indicates more heterogeneity between strong and average teams within the Women's Champions League as compared to Men's. This difference however changes drastically the relaxation timescales of the extreme statistics, as revealed in Fig.~\ref{fig:7}. The distribution of extreme-value timescales peaks close to the first half of the match both for Women's ($21$min) and Men's ($29$min) soccer.  On the other hand, their largest timescale for comebacks differ strongly, $\tau_{\infty}\simeq 61$min for Women's whereas $\tau_{\infty}\simeq 280$min for Men's. For Women's soccer matches one that therefore does not expect a comeback of the weaker after about 60 minutes, whereas for Men's soccer such a comeback is possible until the end of the game. Indeed, only 50\% of the relaxation timescales in Men's soccer can take place in a match of 90min (Fig.~\ref{fig:7} inset). This suggests that shorter matches of less than 70min for Women's soccer would remain   exciting  until the final whistle blows, because the weaker team would yet have chances to win the match.

 It is interesting to compare soccer to other sports such as  handball and basketball, see Fig.~\ref{fig:8}. We estimate the parameters $k_+$ and $k_-$ in a similar manner. For both handball and basketball, men and women, the scoring affinities are close to $A \simeq 0.14$. The scoring rates $\nu$ are given in the caption of Fig.~\ref{fig:8}, in which we can appreciate the difference compared to soccer. For handball and basketball the larger tails towards shorter timescales $\tau_0$ in the Mar{\v{c}}enko-Pastur distributions indicate that the first action or comebacks can happen much more rapidly  than for soccer. 
Handball, basketball and mens soccer share a game duration which falls within the relaxation time spectrum. The long timescale $\tau_\infty$  exceeds significantly the game duration, for all cases except women's soccer.

   \section{Discussion}
   \label{sec5}

We have derived analytical expressions for the  distribution and moments of the finite-time minimum and maximum values of continuous-time biased random walks. Such stochastic processes provide minimal models to describe the fluctuating motion of   molecular motors and cyclic enzymatic reactions that take place in  a thermal reservoir and under non-equilibrium conditions induced   by e.g. external forces and/or  chemical reactions.

Our key results are: (i) exact statistics of the extrema of the position and the entropy production of a biased random walk (ii) a novel connection between extreme-value statistics of biased random walks and the Mar{\v{c}}enko-Pastur distribution of random matrix theory; (iii) symmetry relations between distributions of extrema of stochastic entropy production; (iv) estimates of extreme value statistics from spectral properties of random matrices.
%The key results of our work are specific to one-dimensional biased random walks and stochastic processes described by the combination of $n>1$ independent biased random walks along orthogonal directions. They comprise: (i) distributions and moments of  the finite-time minimum and maximum of the position of the walker and entropy production associated with a stochastic trajectory; (ii) symmetry relations between distributions of finite-time extrema of stochastic entropy production; (iii)  a novel connection between extreme-value statistics of  stochastic processes and the Mar{\v{c}}enko-Pastur distribution from random matrix theory; (iv)  efficient estimates of extreme-value statistics from spectral properties of suitable random matrices.
  
For biased random walks, our results provide insights beyond the infimum law for nonequilibrium steady states, $\langle S_{\rm min}(t) \rangle \geq -1$, which states that the entropy production of a mesoscopic system plus its environment cannot be reduced on average by more than the Boltzmann constant. 
 For  continuous systems  this bound is approached at large times, $\lim_{t\rightarrow \infty}\langle S_{\rm min}(t)\rangle= -1$. Here we have shown that the effects of discreteness are 
very important.  At large times we find that a model dependent bound above $-1$ is reached,
see Eq. (4) and Fig. 2. 
Equations~\eqref{eq:symmean} and~\eqref{eq:symdist} reveal that, for this class of models, the maximum difference between the entropy production and its initial value has the same statistics as the maximum difference between its maximum and its final value, for any given time interval $[0,t]$. Moreover, this result reveals that a "supremum law" for entropy production bounds the average of the difference between entropy production's maximum and its value at a fixed time $t\geq 0$, i.e. 
\begin{equation}
\langle S_{\rm max}(t)\rangle - \langle S(t)\rangle\leq 1\quad.
\label{eq:suplaw}
\end{equation}
The inequality~\eqref{eq:suplaw} follows applying the results of Sec.~V in  Ref.~\cite{neri2017statistics} to the process  $R(t)=\mathcal{P}(X_{[0,t]})/\mathcal{P}(\tilde{X}_{[0,t]})$, which is a martingale with respect to the time-reversed measure.    We have shown that the inequalities for the average  extrema of entropy production and displacement of a biased random walk saturate in the limit of small affinity $A\ll 1$. This limit corresponds to systems that exchange a small amount of heat with their environment --- below the thermal energy $k_{\rm B}T$ --- in each forward or backward step of the walker. For larger values of $A$, our analytical expressions reveal that  details on the discreteness in the walker's motion have strong influence in the extrema statistics. At large times, the time-asymmetric parameter  $A$ fully determines the distribution of extrema, whereas extrema at finite times are determined by both $A$ and the time-symmetric rate constant $\nu$, see Eq.~\eqref{eq:2}.  Moreover, our results for 1D biased random walks also apply to homogeneous 2D biased random walks  describing the motion of molecular motors, upon suitable definitions of effective affinity and rate parameters given by Eqs.~\eqref{eq:nueff} and~\eqref{eq:Aeff}.

We have also revealed a connection between the statistics of finite-time extrema of biased random walks with random-matrix theory, Eqs.~\eqref{eq:MP} and~\eqref{eq:GFext}. We show that the average of finite-time extrema (minimum and maximum) of a 1D biased random walk has a relaxation time spectrum given by Mar{\v{c}}enko-Pastur distribution of eigenvalues of a random matrix ensemble, see Eq.~\eqref{eq:MPdefMT}. 
 Therefore, efficient estimates for the extreme-value statistics of biased random walks can be obtained from the eigenvalue distributions of suitable  Wishart and Laguerre random matrices~\cite{nica2006lectures,fridman2012measuring,livan2018introduction}. Our numerical simulations show that random-matrix estimates can overperform the accuracy and convergence of Monte Carlo simulations to determine extrema statistics.  We remark that the connection between extrema and random-matrix theory developed here is of distinct nature to the relation between statistical properties of stochastic  processes (e.g. interface growth) and  extreme  eigenvalues of random matrices described by a  Tracy-Widom distribution~\cite{tracy1994level,tracy1996orthogonal,prahofer2000universal,dhar2017exact}. 
We expect that this surprising link of stochastic thermodynamics with random matrix theory may turn out to be more general, through the universal convergence of Wishart-Laguerre random matrices in the Marchenko-Pastur theorem, and could hold for larger classes of systems in a similar way.
   
Our work has important consequences for the theory of nonequilibrium fluctuations of active molecular processes and biomolecules.  For example, the statistics of the maximum excursion of a  motor against its net motion  along a track provides insights on the physical limits of pernicious effects of fluctuations at finite times, which can be relevant in e.g. the finite-time efficiency of enzymatic reactions responsible of polymerization processes, muscle contraction by molecular motors, etc. We have shown that the displacement of motors with small cycle affinity  exhibits large extreme values on average. However, the associated extreme entropy flows are  on average always bounded in absolute value by the Boltzmann constant.    Insights of our theory could be also  discussed in the context of more complex biomolecular stochastic processes (e.g. microtubule growth~\cite{helenius2006depolymerizing,redemann2017c} and transport in actin networks~\cite{richard2019active}).  It will be interesting to extend our theory  to   Markovian  and non-Markovian processes with time-dependent driving~\cite{harris2007fluctuation,strasberg2016nonequilibrium,harris2019thermodynamic,proesmans2019phase,loos2019heat,oberreiter2019subharmonic}, stochastic processes with hidden degrees of freedom~\cite{polettini2017effective,martinez2019inferring}, and also to explore whether extrema statistics from single-molecule experimental data    reveal  relaxation spectra described by the Mar{\v{c}}enko-Pastur distribution.

We have illustrated our results in  simple stochastic models of sports (soccer, handball and basketball) yielding the following conclusions. The Mar{\v{c}}enko-Pastur distributions of the extreme-value relaxation times gives an interesting representation of the thrill of the game across timescales, which is both dependent on the rate $\nu$, quite characteristic of the sport, and the strength difference $A$ between the teams. The duration of the game acts as a cursor within this distribution that balances the predictability versus the randomness of the game, favoring the possibility for the weaker team to win from a great coup. Women's soccer appears as an outlier in the comparison of several champions leagues for different sports and genders, that makes the outcome of the game more deterministic. In order to guarantee exciting games, women's soccer matches duration may thus be optimized to shorter durations (60 to 75min, based on the women to men ratio of $\tau_0$ or $\tau_1$) without affecting the comeback possibilities.
 
 We acknowledge fruitful discussions with Izaak Neri, Simone Pigolotti, Ken Sekimoto,  Carlos Mej\'ia-Monasterio, and enlightening discussions with Pierpaolo Vivo. AG acknowledges MPIPKS and ICTP for their hospitality, and Bertrand Fourcade for his guidance towards MPIPKS, and EUR Light S\&T for funding.

\appendix
\section*{Appendix}

\section{First-passage-times, large deviations, and absorptions probabilities of  biased random walks}
\label{app:a}

In this section we review some knowledge of random walk theory (e.g. first-passage statistics~\cite{redner2001guide}) to derive Eq.~\eqref{fptd1} in the Main Text i.e. the exact formula for the first-passage time distribution of a 1D continuous-time biased random walk. We also discuss large-deviation properties of this model, absorption probabilities using martingale theory, as well as parameters estimation from observed trajectories.

\subsection{Model and solution of the Master equation}

We consider a continuous-time biased random walk in a discrete one-dimensional lattice, where $X(t)=\{ 0, \pm 1, \pm 2,\dots \}$ denotes the position of the walker at time $t \geq 0$. We assume that $X(0)=0$ and that the walker can jump from state $X(t)=x$ to $x+1\, (x-1)$ at a rate  $k_+\, (k_-)$.

The waiting time at any site is exponentially distributed with the rate parameter $k_++k_-$. 
The probability $P_x(t)=P\left(X(t)=x\right)$ to find the walker at the lattice site $x$ at time $t$ obeys the master equation
\begin{equation}\label{eq:s1}
\frac{\d P_x(t)}{\d t}=k_+P_{x-1}(t)-(k_+ +k_-)P_x(t)+k_-P_{x+1}(t)\quad,
\end{equation}
with initial condition $P_x(0)=\delta_{x,0}$. From this evolution equation, a velocity $v=k_+-k_-$ and a diffusion coefficient $D=(k_++k_-)/2$ can be defined.
Its solution is the Skellam distribution~\cite{skellam1946frequency} $P_{\rm Sk}(x;\mu_1,\mu_2) = (\mu_1/\mu_2)^{x/2} I_x(2\sqrt{\mu_1\mu_2}) e^{-(\mu_1+\mu_2)} $ with parameters $\mu_1=k_+t$ and $\mu_2=k_-t$
\begin{equation}\label{eq:s2}
P_x(t)=\left(\frac{k_+}{k_-}\right)^{x/2}I_x\left(2\sqrt{k_+k_-}\, t\right)e^{-(k_++k_-) t}\quad,
 \end{equation}
where $I_x(y)$ denotes the $x-$th order modiffied Bessel function of the first kind. Equation~\eqref{eq:s2}  follows from~\eqref{eq:s1} and the exact expression for the generating function of the modified Bessel function of the first kind $\sum_{x=-\infty}^{\infty}z^x I_x(y)=e^{y(z+z^{-1})/2}$. 

%  {\color{blue}
%\subsection{Inferring parameters from observed trajectories}
%
%For a given stochastic trajectory $X(t)$, the parameters describing the stochastic behaviors of the
%one dimensional drift diffusion process can be estimated as follows.
%the rates $k_\pm$ are related to the observables $\langle X(t) \rangle = (k_+-k_-)t$ and $\textrm{Var}[X(t)] = (k_++k_-)t$.
%This implies 
%\begin{align}
%A&=2\tanh^{-1}\frac{\langle X \rangle}{\textrm{Var}[X]} \\
%2\nu &=\frac{\langle X(t)\rangle}{t}\sqrt{\left(\frac{\textrm{Var}[X]}{\langle X\rangle}\right)^2-1} \quad,
%\end{align}
%as well as
%\begin{align}
%\langle X_{\min}\rangle=-\frac{1}{2}\left(\frac{\textrm{Var}[X]}{\langle X \rangle}-1\right)
%\end{align}
%when the bias is positive. Note that the average velocity $\frac{\langle X(t)\rangle}{t}$ and the Fano factor $\frac{\textrm{Var}[X]}{\langle X \rangle}$, that appears naturally here, can be measured in a robust way, even when the measured trajectories are very noisy~\cite{svoboda1994fluctuation}. The above expressions thus provide a direct estimate of the extreme-value statistics for these processes.
%
%}

\subsection{Large deviation and diffusion limit}

We now discuss and review  large deviation properties of the 1D biased random walk~\cite{touchette2009large,speck2012large} and relate them to the statistics of 1D drift diffusion process.  For this purpose we consider the scaling limit of $P_x(t)$ given by Eq.~\eqref{eq:s2} for large $x\sim vt$, with $v=k_+-k_-$ the net velocity of the walker. We assume  a large deviation principle for $P_x(t)$ of the form
\begin{equation}
P_x(t)\sim e^{-2\nu t J(x/v t)}\quad.
\end{equation}
In order to derive an analytical expression for the rate function $J$, we first approximate 
 the modified Bessel function in Eq.~\eqref{eq:s2}  using a saddle point approximation 
\begin{eqnarray}
I_x(x/z)&=&\frac{1}{2\pi}\int_0^{2\pi} e^{x(z^{-1}\cos \theta -i\theta)}\text{d}\theta \\
&\sim& \frac{e^{x(z^{-1}\cos \theta_0 -i\theta_0)}}{\sqrt{2\pi xz^{-1}\cos \theta_0}} = \frac{e^{x\left(\sqrt{1+z^{-2}}-\sinh^{-1}z\right)}}{\sqrt{2\pi x\sqrt{1+z^{-2}}}} \quad, \label{eq:sdlpt}
\end{eqnarray}
 where the saddle point is given by $i\sin \theta_0=z$ i.e. $i\theta_0=\sinh^{-1}z$ and we have used $\cos i\theta_0=\cosh \theta_0$ and $z^{-1} \cosh \sinh^{-1} z=\sqrt{1+z^{-2}}$. The rate function $J(u)$ with $u=x/v t$ can be evaluated from the leading term of  Eq.~\eqref{eq:s2}   which is found using Eq.~\eqref{eq:sdlpt}: 
\begin{eqnarray}
\hspace{-1cm}J(u)&=&J\left(z/\sinh(A/2)\right) \nonumber\\
&\equiv&\lim_{t \to \infty}-\frac{\ln P_{2 \nu t z}(t)}{2\nu t}\nonumber \\
&=&\lim_{t \to \infty}\frac{2\cosh(A/2)\nu t  -\nu t z A-\ln I_{2 \nu t z}(2\nu t)}{2\nu t}  \nonumber \\
&=&\cosh\frac{A}{2}+z\left(\sinh^{-1}z-\frac{A}{2}\right)-\sqrt{1+z^2} \quad,
\end{eqnarray}
where $z=u\sinh(A/2)=x/2\nu t$ and the change of variables $(k_{+},k_-)\to (A,\nu)$ have been used for convenience.

In the vicinity of the minimum where $u\sim 1$ and thus $z\sim \sinh(A/2)$, the large deviation function behaves as
\begin{equation}
J(u)=\frac{\sinh(A/2)^2(u-1)^2}{2\cosh(A/2) }+\frac{\sinh(A/2)^4(u-1)^3}{6\cosh(A/2)^3 }+O(u-1)^4. \label{eq:a77}
\end{equation}
Interestingly, the ratio between the second  and the leading term of~\eqref{eq:a77}, given by $\tanh(A/2)^2(u-1)/3$,  vanishes for small deviations $u\sim1$ but also for large deviations in the limit of a small bias $A\ll1$. The continuum limit of the biased random walk  for $A$ small simplifies to the drifted Brownian motion $P_x(t)=e^{-(x-vt)^2/4Dt}/\sqrt{4\pi Dt}$, 
where the polynomial prefactor is recovered by normalization, and we have used the expressions for the velocity $v=2\nu\sinh(A/2)$ and the diffusion coefficient $D=\nu\cosh(A/2)$. In this regard, the 1D biased random walk can be seen as a generalization of the drifted Brownian motion for any bias. A finite bias modifies occurrences of  extreme large deviations with respect to those occurring in the drift diffusion process.  Consequently, the bias $A$  is expected to affect the extreme value statistics of the process, as shown below.

\subsection{Martingales and absorption probabilities}
In this subsection we employ martingale theory to derive an analytical expression of the absorption probability $P_{\rm abs}(-x)$ for a 1D biased random walk starting at $x=0$ to ever reach an absorbing boundary located at $-x<0$.

We first show explicitly that $e^{-S(t)}$ is a martingale process with respect to $X(t)$, i.e.
\begin{equation}
\langle e^{-S(t)} | X_{[0,t']} \rangle  = e^{-S(t')} \quad, \label{eq:mart}
\end{equation}
for $t\geq t'$. In words, the average of $e^{-S(t)}$ over all trajectories with common history $X_{[0,t']}$ up to time $t'\leq t$ equals to its value at the last time of the conditioning $e^{-S(t')}$. The proof is as follows:
\begin{eqnarray}
\hspace{-0.5cm}\langle e^{-S(t)} | X_{[0,t']} \rangle  &=& \langle  e^{-[S(t)-S(t')]}  | X_{[0,t']} \rangle \, e^{-S(t')}  \nonumber\\
&=& \langle  e^{-S(t-t')}  \rangle \, e^{-S(t')} \nonumber\\
&=& e^{-S(t')} \sum_{x=-\infty}^{\infty} P_x(\Delta t) \left(\frac{k_-}{k_+}\right)^{x}\nonumber\\
&=& e^{-S(t')} \, e^{-(k_++k_-) \Delta t} \nonumber\\
&& \times \underbrace{\sum_{x=-\infty}^{\infty}\left(\frac{k_+}{k_-}\right)^{-x/2}I_x\left(2\sqrt{k_+k_-}\, \Delta t\right)}_{= e^{(k_++k_-) \Delta t}} \nonumber\\
&=& e^{-S(t')} \quad.\quad\square
\end{eqnarray}
In the first and second lines we have used  the additive property and the Markov property of entropy production, respectively. In the third line we have used the definitions $\Delta t\equiv t-t'$ and $S(t) = X(t)\ln(k_+/k_-)$. In the fourth line we have used the identity $\sum_{x=-\infty}^{\infty}z^x I_x(y)=e^{y(z+z^{-1})/2}$.

 We remark that the proof sketched above can be simplified using the integral fluctuation relation $\langle e^{-S(t-t')}\rangle=1$ in  the second line, which holds for any $t\geq t'$~\cite{jarzynski1997nonequilibrium,seifert2005entropy}. It has been  shown~\cite{neri2019integral,neri2017statistics} that the martingality of $e^{-S(t)}$ implies  a set of integral fluctuation relations at stopping times
\begin{equation}
\langle e^{-S(\T)} \rangle=1\quad, \label{eq:IFTST}
\end{equation}
where $\T$ is any bounded stopping time, i.e. a stochastic time at which the process $X(t)$ satisfies for the first time a certain criterion. In particular, Eq.~\eqref{eq:IFTST} holds for the first-passage time $\T_2$ of $X(t)$ to reach any of two absorbing barriers located at $-x_-$ and $x_+$, with $x_+$ and $x_-$ two arbitrary positive integer numbers. When applying Eq.~\eqref{eq:IFTST}  to this particular stopping time, we can unfold the average in the left-hand side using the absorption probabilities
\begin{eqnarray}
\langle e^{-S(\T_2)} \rangle &=& P_{\rm abs}(x_+) \langle e^{-S(\T_2)}\rangle_+  + P_{\rm abs}(x_-) \langle e^{-S(\T_2)}\rangle_- \nonumber\\
&=& P_{\rm abs}(x_+) e^{-Ax_+} + P_{\rm abs}(x_-) e^{Ax_-} \nonumber\\
&=& e^{-Ax_+} + P_{\rm abs}(x_-) [ e^{Ax_-} -  e^{-Ax_+}] \nonumber\\
&=&1\quad, \label{eq:st2}
\end{eqnarray}
where in the second line we have used the fact that $e^{-S(\T_2)} = e^{-A x_+} $ with probability $P_{\rm abs}(x_+)$ and $e^{-S(\T_2)} = e^{A x_-} $ with probability $P_{\rm abs}(x_-)$. In the third line we have used $P_{\rm abs}(x_+) +P_{\rm abs}(x_-) =1$, and in the fourth line Eq.~\eqref{eq:IFTST}. Solving  the third line Eq.~\eqref{eq:st2} for the absorption probability we obtain
\begin{equation}
P_{\rm abs}(x_-) = \frac{1- e^{-Ax_+}}{e^{Ax_-} -  e^{-Ax_+}}\quad. \label{Pminusint}
\end{equation}
Taking the limit $x_+\to \infty$ in Eq.~\eqref{Pminusint} we obtain the well-known analytical expression for the absorption probability 
\begin{equation}
P_{\rm abs}(x) = e^{-Ax}\quad,
\end{equation}
which we used to derive the analytical expressions $-$ Eqs.~\eqref{eq:PSmin} and~\eqref{eq:Smin} $-$ of the distribution and mean of the global minimum of entropy production in the biased random walk.

\subsection{First-passage-time distribution}

The first-passage-time density  $P_\textrm{fpt}(t';x)$ can be derived from the solution of  the Master equation~\eqref{eq:s1} with an absorbing boundary at site $x\neq 0$, with $x$ an integer number~\cite{redner2001guide}. It can also be derived from the distribution of the walker using Laplace transforms through the renewal equation:
 \begin{align}
P_x(t)=\int_0^t P_\textrm{fpt}(t';x)P_0(t-t')\d t' \quad .
 \end{align}
 where $P_0(t)$ is  the probability to be at a state at time $t$ when the system was at the same state at $t=0$.
This convolution integral becomes a product in the Laplace domain, for any $x\neq 0$:
 \begin{equation}\label{eq:s4}
\hat{P}_x(s)=\hat{P}_0(s)\hat{ P}_\textrm{fpt}(s;x) =\frac{e^{Ax/2} \left(h(s)+\sqrt{h(s)^2-1}\right)^{-x}}{2 \nu \sqrt{h(s)^2-1}}\quad,
\end{equation}
where 
\begin{equation}\label{eq:s5}
 h(s)\equiv \frac{s+k_++k_-}{2\sqrt{k_+k_-}}=\frac{s}{2\nu}+\cosh (A/2)\quad,
 \end{equation}
 and 
 \begin{equation}\label{eq:s6}
 \hat{P}_0(s) = \int_0^{\infty}\text{d}t \, e^{-st}P_0(t)=\frac{1}{2\nu \sqrt{h^2(s)-1}}\quad.
 \end{equation}
 We thus obtain, using Eqs.~\eqref{eq:s5} and~\eqref{eq:s6} in~\eqref{eq:s4}
 \begin{eqnarray}\label{eq:s7}
\hat{P}_\textrm{fpt}(s;x)&=&e^{Ax/2} \left(h(s)+\sqrt{h(s)^2-1}\right)^{-x}\nonumber\\
&=&e^{Ax/2-|x|\cosh^{-1}\left(s/2\nu+\cosh (A/2)\right)} \quad. \label{fptLaplace}
 \end{eqnarray}
 In the above  equations 
 and in the following we will use the variables 
 $\nu=(k_+ k_-)^{1/2}$ and 
 $A=\ln(k_+/k_-)$, see Eq.~\eqref{eq:2} in the Main Text. The inverse Laplace transform of Eq.~\eqref{eq:s7} 
implies  $P_{\textrm{fpt}}(t;x)=(|x|/t)P_x(t)$:
\begin{equation}
P_{\textrm{fpt}}(t;x)=\frac{|x|}{t}e^{Ax/2} I_x(2\nu t)e^{-2\cosh(A/2)\nu t}  \quad ,\qquad\square
\label{eq:fptd}
\end{equation}
which is Eq.~\eqref{fptd1} in the Main Text.

\section{ Exact  extrema statistics\\  for the  1D   biased random walk}
\label{app:b}

In this section, we use generating functions to derive the statistics of the finite-time extrema of entropy production. In particular, we focus on the  generating function for the probability $G_{\min}$ of the minimum  and the generating function for the absorption probability $G_{\rm abs}$, which are defined respectively as
\begin{eqnarray}
G_{\min}(z;t)&\equiv&\sum_{x= 0}^{\infty} z^{-x} P\left(S_{\min}(t)=-A x\right)\quad,\label{eq:GG}\\
G_{\rm abs}(z;t)&\equiv&\sum_{x=1}^{\infty} z^{-x} P_{\rm abs}(-x;t)\quad.
\end{eqnarray}
These two generating functions are related by
\begin{equation}
G_{\min}(z;t)=1+G_{\rm abs}(z;t)(1-z) \label{relationGG}\quad .
\end{equation}
Moments and probabilities follow taking derivatives of the generating functions~\eqref{eq:GG} with respect to $z$
\begin{align}
\langle S_{\min}(t)^m \rangle &= \left.\frac{\partial^m G_{\min}(z^A;t)}{\partial (\ln z )^m}\right\rvert_{z=1}\quad, \label{Moments}\\
 P\left(S_{\min}(t)=-mA\right) &= \left.\frac{\partial^m G_{\min}(z;t)}{m!\;\partial (z^{-1})^m}\right\rvert_{z^{-1}=0} \label{Proba}\quad .
\end{align}
For instance, inserting~\eqref{relationGG} into~\eqref{Moments} to compute the first moment reduces to the simple expression
\begin{equation}\label{eq:gfmin}
\langle S_{\min}(t) \rangle=-AG_{\rm abs}(1;t)\quad.
\end{equation}

\subsection{Finite time statistics in the Laplace domain}

We rewrite the probability-generating function $G_{\min}$ using the first-passage-time density derived above~\eqref{fptLaplace}. First we use the fact the the first-passage-time density is the derivative of the absorption probability,
$\hat{P}_\textrm{fpt}(s;x)=s\hat{P}_\textrm{abs}(x;s)-\delta_{x,0}$, which holds for any $x$. Equation~\eqref{eq:fptd} implies that the Laplace transform $\hat{P}_\textrm{fpt}(s;-x)$ of the first-passage-time probability to reach an absorbing site located at $-x$, with $x\geq 1$, can be expressed  as the $x-$th power of the Laplace transform $\hat{P}_\textrm{fpt}(s;-1)$ of the first-passage time probability to reach $x=-1$:
\begin{equation}\label{eq:powerfpt}
\hat{P}_\textrm{fpt}(s;-x)=\hat{P}_\textrm{fpt}(s;-1)^{x} \quad .
\end{equation}
Consequently, the statistics of the minimum (maximum) can be expressed in terms of just  $\hat{P}_\textrm{fpt}(s;-1)$ ($\hat{P}_\textrm{fpt}(s;1)$). Using Eqs.~(\ref{eq:GG}-\ref{relationGG}) and~\eqref{eq:powerfpt}, we derive the generating functions in terms of $\hat{P}_\textrm{fpt}(s;-1)$
\begin{align}
s\hat{G}_{\rm abs}(z;s)&=\sum_{x=1}^{\infty} z^{-x}\hat{P}_\textrm{fpt}(s;-1)^{x}=\frac{z^{-1}\hat{P}_\textrm{fpt}(s;-1)}{1-z^{-1}\hat{P}_\textrm{fpt}(s;-1)}\;, \label{eq:gflaplace}\\
s\hat{G}_{\min}(z;s)&=\frac{1-\hat{P}_\textrm{fpt}(s;-1)}{1-z^{-1}\hat{P}_\textrm{fpt}(s;-1)} \quad .
\end{align}
Because $\hat{P}_\textrm{fpt}(s;-1)=e^{-A/2} (h(s)+\sqrt{h(s)^2-1})$ is algebraic, all the moments and probabilities, obtained from Eqs.~(\ref{Moments}-\ref{Proba}), are algebraic expressions. For instance, the Laplace transform of the mean minimum is obtained directly from~\eqref{eq:gfmin} and~\eqref{eq:gflaplace}:
\begin{align}\label{eq:s182}
s\langle \hat{S}_{\min}(s) \rangle=\frac{-A}{\hat{P}_\textrm{fpt}(s;-1)^{-1}-1} \quad.
\end{align}

\iffalse
It may be desirable to rewrite Laplace domain quantities as $\frac{1}{\lambda+s}$ multiplied by a weight for each possible lambda. Indeed, each of these terms is easily inverted as a decaying exponential $e^{-\lambda t}$. This translation of the inverse Laplace transform of a quantity as the Laplace transform of another quantity can be view as a transformation of one quantity into the other, that is called the Cauchy-Stieltjes transform:
\begin{align}
\hat{g}(s)&=\int \frac{\rho(\lambda)}{\lambda+s}\d \lambda\\
g(t)&=\int \rho(\lambda)e^{-\lambda t}\d \lambda
\end{align}
\fi

\subsection{Integral representations of extreme value statistics}

We start from the first-passage-time density formula~\eqref{eq:fptd} and we exploit two properties of the modified Bessel function of the first kind. This allows us to rewrite the absorption probability as a definite integral of trigonometric and hyperbolic functions and the parameters $A$ and $\nu$:
\begin{eqnarray}
&&\hspace{-1cm}P_\textrm{abs}(x;t) =\int_0^t P_\textrm{fpt}(t';x)\d t'\nonumber\\
&&\hspace{-1cm}=e^{Ax/2}\int_0^t\frac{|x|}{t'}I_x(2\nu t')e^{- 2\cosh(A/2)\nu t'}\d t' \label{eq:s17}\\
&&\hspace{-1cm}=e^{Ax/2}\int_0^{\pi} \int_0^{t} e^{-2\nu t' (\cosh(A/2)-\cos \theta)} \nonumber\\
&&\hspace{-0.5cm}\times 2\nu\d t'\left(\cos\left[(|x|-1)\theta\right]-\cos\left[(|x|+1)\theta\right]\right) \frac{\d\theta}{2\pi} \label{eq:s18}\\
&&\hspace{-1cm}= e^{Ax/2}\int_0^{\pi}\frac{1-e^{-2\nu t (\cosh(A/2)-\cos \theta)}}{\cosh(A/2)-\cos \theta}\nonumber\\
&\times&\left(\cos\left[(|x|-1)\theta\right]-\cos\left[(|x|+1)\theta\right]\right) \frac{\d\theta}{2\pi} \quad  .\label{eq:s19}
\end{eqnarray}
In~\eqref{eq:s17} we have used Eq.~\eqref{eq:fptd}. In~\eqref{eq:s18} we have used the definition $I_x(y) = (1/\pi)\int_0^{\pi} e^{y\cos \theta }\cos(x\theta)\d \theta$ and the property  $I_x(y)=(y/2x)[I_{x-1}(y)-I_{x+1}(y)]$. Finally in~\eqref{eq:s19} we have performed the integration over $t$. 
Using Eq.~\eqref{eq:s19}, we  express the generating function of the absorption probability as an integral:
\begin{eqnarray}
&& G_{\rm abs}(z;t)=\sum_{x=1}^{\infty} z^{-x} P_{\rm abs}(-x;t) \label{eq:s21}\\
&&=\int_0^\pi \frac{1-e^{-2\nu t(\cosh(A/2)-\cos \theta)}}{\cosh(A/2)-\cos \theta}\frac{(\sin \theta)^2}{\cosh(A/2+\ln z)-\cos \theta}\frac{\d\theta}{2\pi},\nonumber
\end{eqnarray} 
where we have used in the above equation the identity
\begin{equation}
\sum_{x=1}^{\infty}e^{-\alpha x}\Big(\cos\left((x-1)\theta\right)-\cos\left((x+1)\theta\right)\Big)=\frac{(\sin \theta)^2}{\cosh\alpha-\cos \theta},
\end{equation}
which follows from the generating function of Chebyshev polynomials of the first kind $T_x(\cos \theta)\equiv \cos (x\theta)$.
Performing  the change of variable $y=\cos \theta $ in Eq.~\eqref{eq:s21}, and setting $z=1$ (we recall $\langle S_{\min}(t) \rangle=-AG_{\rm abs}(1,t)$, see Eq.~\eqref{eq:gfmin}) we obtain Eq.~\eqref{eq:integral} in the Main Text:
\begin{eqnarray}
\langle S_{\min}(t) \rangle &=& -\frac{A}{2\pi}\int_{-1}^1 \frac{1-e^{-2\nu t\left(\cosh(A/2)-y\right)}}{\left(\cosh(A/2)-y\right)^2}\sqrt{1-y^2}\d y  . \;\square\nonumber\\
 \label{intrpz}
\end{eqnarray}

\section{Extrema statistics and  Mar{\v{c}}enko-Pastur distribution}
\label{app:c}

In this Section, we derive analytical expressions for the extrema statistics of 1D biased random walks in terms of the Mar{\v{c}}enko-Pastur 
 distribution~\eqref{eq:MPdefMT}  of random matrix theory, copied here for convenience:
\begin{equation}
\rho(\lambda)\equiv 
 \begin{cases} \dfrac{1}{2\pi \delta}\dfrac{\sqrt{(\lambda_+-\lambda)(\lambda-\lambda_-)}}{\lambda} &\;\mbox{if}\;\; \lambda \in [\lambda_-,\lambda_+]\quad \\
0 & \;\mbox{if}\;\;\lambda \notin [\lambda_-,\lambda_+],\quad \end{cases} \label{eq:MPdef}
  \end{equation}
 where $\lambda$ is a positive random variable, $\delta\leq1$ a parameter, and $\lambda_\pm=\left(1\pm\sqrt{\delta}\right)^2$. Note that this distribution is normalized 
 with mean $\int_{\lambda_-}^{\lambda_+} \lambda\rho(\lambda)\text{d}\lambda=1$.

\subsection{Average value of the finite-time minimum of entropy production}

Performing the changes of variable $k = 2\nu (\cosh(A/2)-y)$, as well as $\tau=1/k$, in  Eq.~\eqref{intrpz} we obtain
 \begin{align}
\langle S_{\min}(t) \rangle&=-\frac{A}{2\pi}\int_{k_0}^{k_\infty}\frac{1-e^{-kt}}{k}\sqrt{(k_\infty-k)(k-k_0)}\frac{\d k}{k}\label{kform} \\
 &=-\frac{A}{2\pi}\int_{\tau_0}^{\tau_\infty} \left(1-e^{-t/\tau}\right)\sqrt{\frac{(\tau_\infty-\tau)(\tau-\tau_0)}{\tau_\infty \tau_0}}\frac{\d \tau}{\tau} \label{tauform}
\quad 
\end{align}
where we have introduced the variables
\begin{eqnarray}
k_{\infty/0} &\equiv& (\sqrt{k_+} \pm \sqrt{k_-})^{2}\quad,\\ 
\tau_{\infty/0} &\equiv& \frac{1}{k_{0/\infty}}\quad.
\end{eqnarray}
 Thus, the average entropy production minimum can be  expressed as exponential relaxation process with 
a spectrum of relaxation times  distributed according to Mar{\v{c}}enko-Pastur distributions~\eqref{eq:MPdef} 
\begin{eqnarray}
\langle S_{\min}(t) \rangle &=&  -Ak_- \int_{k_0}^{k_\infty} \frac{1-e^{-kt}}{k}\rho(k/\bar{k})\frac{\d k}{\bar{k}}  \label{kform1} \\ 
&=&  \langle S_{\rm min} \rangle \int_{\tau_0}^{\tau_\infty} (1-e^{-t/\tau})\rho(\tau/\bar{\tau})\frac{\d \tau}{\bar{\tau}}  \label{tauform1} \quad,
\end{eqnarray}
 where 
\begin{eqnarray}
\bar{k}&\equiv&k_+\quad,  \label{eq:barr}\\ 
\bar{\tau}&\equiv& \frac{k_+}{(k_+-k_-)^2} =\frac{\sqrt{\tau_0 \tau_\infty}}{1-e^{-A}} \label{eq:bartau}\quad,
\end{eqnarray}
and 
\begin{equation}
\delta=\frac{k_-}{k_+}=e^{-A}\quad.
\end{equation}
Note that  Eq.~\eqref{tauform1} provides Eq.~\eqref{eq:MP} of the Main Text.

\subsection{Generating functions of the absorption probability and the minimum}
 
 The generating function for the absorption probability  can also be expressed using {Mar{\v{c}}enko-Pastur 
distributions. Using the same method as described above for  Eq.~\eqref{eq:s21}, we find
\begin{eqnarray}
G_{\rm abs}(z;t)&=&k_- \int_{k_0}^{k_\infty} \frac{1-e^{-kt}}{k+f(z)} \rho(k/\bar{k})\d k/\bar{k} \label{GintMPk} \\
&=&\frac{1}{e^A-1}\int_{\tau_0}^{\tau_\infty} \frac{1-e^{-t/\tau}}{1+f(z)\tau}\rho(\tau/\bar{\tau})\d \tau/\bar{\tau}\;\;, \qquad\label{GintMPtau}
\end{eqnarray}
where 
\begin{equation}
f(z)\equiv k_+(z-1)+k_-(z^{-1}-1)\quad.
\end{equation} 
Using Eqs.~\eqref{GintMPtau} and~\eqref{relationGG} we obtain the generating function of the distribution of minima given by Eq.~\eqref{eq:GFext} in the Main Text
\begin{equation}
G_{\min}(z;t)=1+\frac{1-z}{e^A-1}\int_{\tau_0}^{\tau_\infty} \frac{1-e^{-t/\tau}}{1+f(z)\tau}\rho(\tau/\bar{\tau})\d \tau/\bar{\tau} \label{relationGG2}\quad .\quad\square
\end{equation}

\subsection{Laplace transforms}
Taking the Laplace transform of Eq.~\eqref{GintMPk}, we obtain
\begin{equation}\label{GintMPLaplace}
s\hat{G}_{\rm abs}(z;s)= k_- \int_{k_0}^{k_\infty} \frac{k}{k+s} \frac{\rho(k/\bar{k})}{k+f(z)}\frac{\d k}{\bar{k}} \quad.
\end{equation}
Equations~\eqref{GintMPLaplace} and~\eqref{eq:gfmin} imply  that the Laplace transform of the average minimum can be written as a Stieltjes-like transform of the Mar{\v{c}}enko-Pastur distribution for the variable $k$
\begin{equation}\label{eq:s34}
s\langle \hat{S}_{\min}(s) \rangle = -Ak_- \int_{k_0}^{k_\infty} \frac{\rho(k/\bar{k})}{k+s} \frac{\d k}{\bar{k}} \quad,
\end{equation}
and similar relations hold for moments of any order. Notably, Eqs.~\eqref{GintMPLaplace} and~\eqref{eq:s34} have a similar mathematical structure as the Laplace transform of the first-passage-time density of  Markovian stochastic processes found in~\cite{hartich2019extreme}, where instead $\hat{P}_{\rm fpt}(s;x)$ is expressed as a weighted discrete sum of relaxation modes.

\section{Random-matrix estimates \\ of extreme-value statistics}
\label{app:d}

In this section we discuss the connection between the relaxation spectrum of first-passage and extrema statistics in the 1D biased random walk with random-matrix theory. 
We now describe how one can estimate finite-time statistics of the minimum entropy production from the spectrum of suitable random matrices.
For this purpose, we use  a  celebrated result by Mar{\v{c}}enko and Pastur~\cite{marvcenko1967distribution}. Consider a  real $m\times m$ Wishart matrix defined as 
\begin{equation}\label{Wishart}
\mathbf{W}=\frac{1}{n}\mathbf{R} \mathbf{R}^T,
\end{equation}
where $\mathbf{R}$ (its transpose $\mathbf{R}^T$) is a  $m\times n$  random matrix, with $n\geq m$.  The random matrix $\mathbf{R}$ is filled with independent identically distributed (i.i.d.)  random variables drawn from a normal distribution of zero mean and unit variance, i.e.  $\mathbf{R}_{ij}\sim \mathcal{N}(0,1)$, for all $i,j\leq m$. The resulting positive definite random matrix follows the Wishart distribution of degree of freedom $n$ and density $c_{n,m}\text{det}\mathbf{w}^{(n-m-1)/2}e^{\Tr \mathbf{w} \,n/2}$ (where $c_{n,m}$ is a normalization factor). Following Mar{\v{c}}enko and Pastur, the eigenvalues $\lambda$ of the Wishart random matrix $\mathbf{W}$ are asymptotically distributed according to the distribution~\eqref{eq:MPdef} in the limit $n,m\to \infty$ with finite \textit{rectangularity} $m/n= \delta<1$. It has been shown that this asymptotic result also holds when all  $\mathbf{R}_{ij}$ are i.i.d. random variables drawn from any distribution of zero mean and unit variance~\cite{livan2018introduction}.

We now put in practice Mar{\v{c}}enko and Pastur's result, namely we find random matrices whose spectral density matches with that of the relaxation spectrum of the average minimum of entropy production. This can be achieved e.g. by using a Wishart random matrix of rectangularity $\delta=m/n$ identified as $k_-/k_+=e^{-A}$ in terms of the bias $A$ of the walker, i.e. we draw a real $m\times n$ Wishart random matrix $\mathbf{W}$ with $m,n\gg1$ and $m/n \simeq  e^{-A}$ (for instance $n=\lceil e^A m\rceil$). 
Then we evaluate the $m$ eigenvalues $\lambda_i$ of the matrix $\mathbf{W}$ and we give them a dimension using Eqs.~(\ref{eq:barr}-\ref{eq:bartau}) and performing the changes of variables $k=\lambda \bar{k}$ in Eq.~\eqref{kform1} and $\tau=\lambda \bar{\tau}$ in Eq.~\eqref{tauform1} respectively. We thus obtain the following two estimators, $\langle \tilde{S}_{\min}(t) \rangle_{k}$ and $\langle \tilde{S}_{\min}(t) \rangle_{\tau}$, for the average entropy production minimum:
\begin{eqnarray}
\langle \tilde{S}_{\min}(t) \rangle_{k} &\equiv& -Ae^{-A} \;\frac{1}{m}\sum_{i=1}^m \frac{1-e^{-\lambda_i \bar{k}t}}{\lambda_i}  \label{RMkest}\quad , \\
\langle \tilde{S}_{\min}(t) \rangle_{\tau}&\equiv&\langle S_{\min}\rangle \;\frac{1}{m}\sum_{i=1}^m  \left(1-e^{-t/\lambda_i \bar{\tau}} \right) \label{RMtauest}\quad ,
\end{eqnarray}
where $\bar{k}=k_+$ and $\bar{\tau}=k_+/(k_+-k_-)^2$ as identified previously. These estimators converge respectively to the exact result in the limit of a large matrix size. 
Using Eq.~(\ref{GintMPk}-\ref{GintMPtau}), the same procedure can be applied to estimate the generating function and any order moment of  the distribution of entropy production extrema.  

To test the convergence of these estimators, we define their relative error $\epsilon_{k}(t)$ and $\epsilon_{\tau}(t)$ as the relative difference
\begin{eqnarray}
\epsilon_{k,\tau}(t)&\equiv& \frac{\langle \tilde{S}_{\rm min}(t)\rangle_{k,\tau}-\langle S_{\rm min}(t)\rangle}{\langle S_{\rm min}(t)\rangle} \quad ,\label{Er_t}
\end{eqnarray}
which is a random real quantity for both $k$ and $\tau$ estimates. Their limiting values are related and can be calculated analytically:
\begin{eqnarray}
\epsilon_{\min}&\equiv&\lim_{t\rightarrow 0} \epsilon_{k}(t)=\lim_{t\rightarrow \infty} \epsilon_{\tau}(t)= 0 \label{Er0} \\
\epsilon_{\max}&\equiv&\lim_{t\rightarrow \infty} \epsilon_{k}(t)=\lim_{t\rightarrow 0} \epsilon_{\tau}(t)\nonumber\\
&=&(1-e^{-A})\left(\frac{1}{m}\sum_{i=1}^m \frac{1}{\lambda_i}\right)-1 \label{Er} \quad,
\end{eqnarray}
which vanishes in the limit of a large random matrix because $\langle 1/\lambda\rangle_{\rho}=1/(1-e^{-A})$,  with $\langle ...\rangle_\rho$ denoting an average over the Mar{\v{c}}enko-Pastur distribution~\eqref{eq:MPdef}}. From the limits~\eqref{Er0}, we conclude that the estimator~\eqref{RMkest} is advantageous to study the short-time behavior whereas~\eqref{RMtauest} is most suited for large-time asymptotics. Our numerical results show that $|\epsilon_{k}(t)| \leq |\epsilon_{\max}|$ and also $|\epsilon_{\tau}(t)| \leq |\epsilon_{\max}|$ for all tested parameter values and for all times $t$. Therefore we will use $\epsilon_{\max}$ given by Eq.~\eqref{Er} as a conservative bound for the relative error of the random-matrix estimates at any time $t$.

\begin{figure*}[ht!]
\includegraphics[width=6cm]{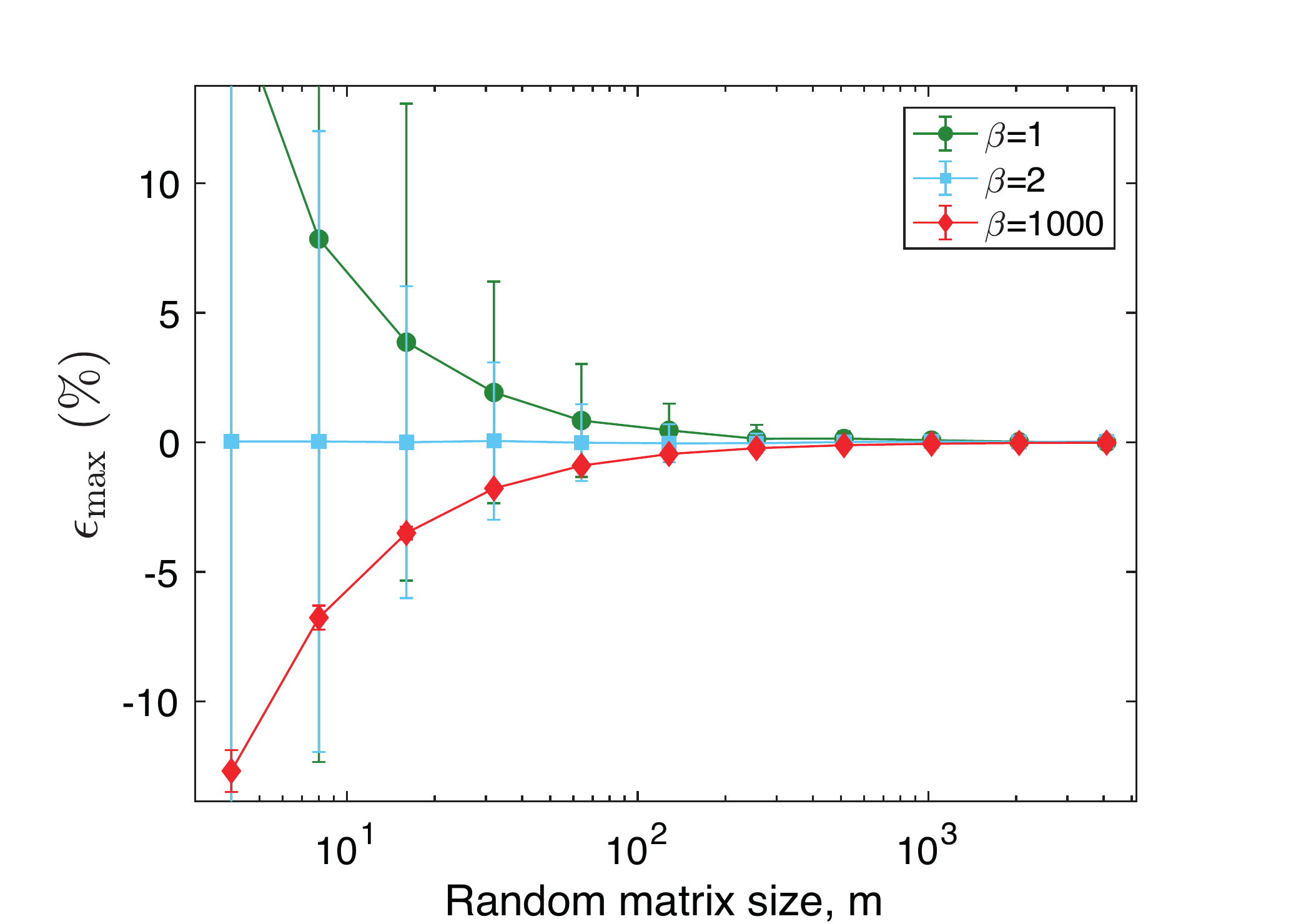}\includegraphics[width=6cm]{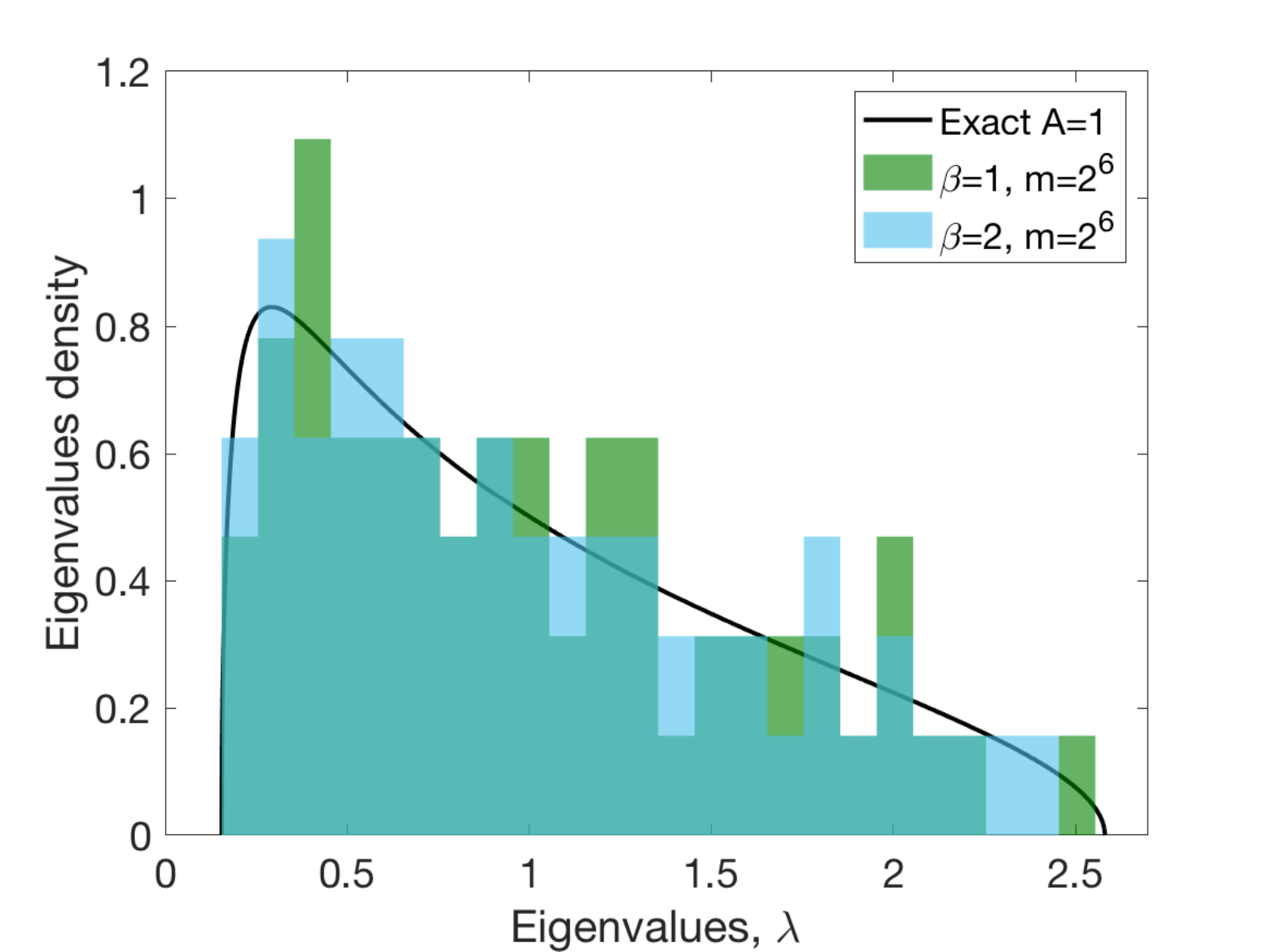}\includegraphics[width=6cm]{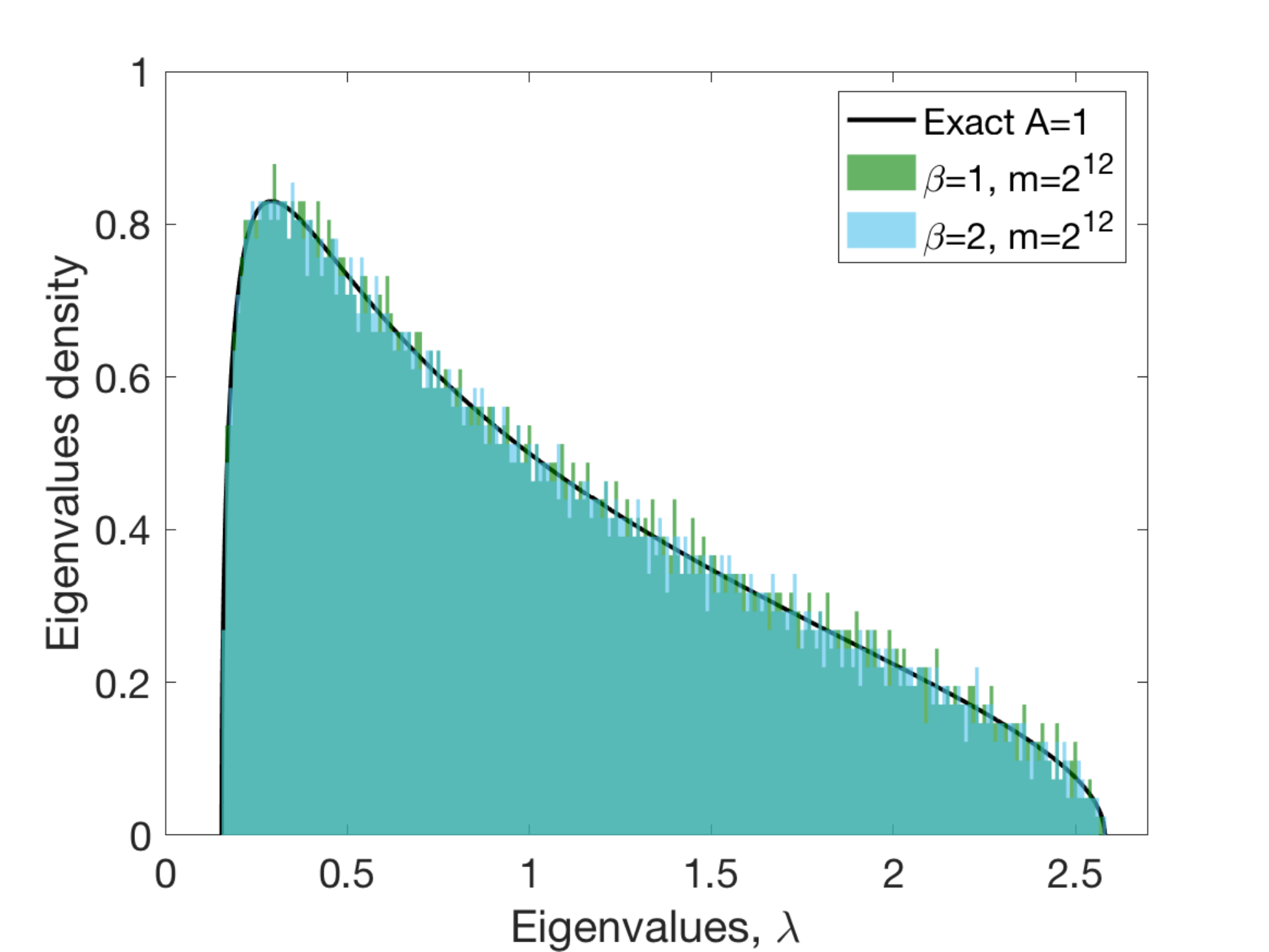}\vspace{0.5cm}
\caption{Numerical results for the random-matrix estimates  $\langle \tilde{S}_{\rm min}(t)\rangle$ obtained using $m\times $m random matrices drawn from the $\beta$-Laguerre ensemble.  Left: Average value of the maximum relative difference $\epsilon_{\rm max}$ given by Eq.~\eqref{Er}  as a function of the  matrix size $m$ obtained from $10^6/m$ random matrices, for three different $\beta-$Laguerre matrix ensembles (see legend). The error bars are given by the standard deviation of $\epsilon_{\rm max}$ across the random-matrix populations. Middle and Right: Normalized histograms for the eigenvalues of a single random matrix drawn from the $\beta$-Laguerre ensemble ($\beta=1,2$, see legend) with matrix sizes $m=2^6$ (middle panel) and $m=2^{12}$ (right panel), compared to the Mar{\v{c}}enko-Pastur distribution~\eqref{eq:MPdef} with parameter $\delta = e^{-A}$ (black line) and  affinity $A=1$.}\vspace{0.5cm}
\label{fig:4s}
\end{figure*}

The estimates introduced above rely on the fact that one can achieve a rectangularity $\delta =m/n\simeq e^{-A}$ with large enough random matrices.  Because $e^{-A}$ is in general a real number, it is desirable to develop random-matrix estimators that achieve the Mar{\v{c}}enko-Pastur distribution accurately for any value of $A$.  Following~\cite{nica2006lectures}, the $\beta$-Laguerre matrices are an alternative ensemble whose spectral density tends asymptotically to the distribution $ \text{MP}(e^{-A})$ in the large size limit. A $\beta$-Laguerre $m\times m$ random matrix  $\mathbf{L}$ is defined as
\begin{equation}\label{Laguerre}
\mathbf{L}=\frac{1}{n}\mathbf{R}\mathbf{R}^{T} \quad , 
\end{equation}
where $n=m\beta/\delta$. Here, $\mathbf{R}$ is a $m\times m$ random matrix with all  entries equal to zero except the $m\times(m-1)$ diagonal and sub-diagonal elements. The non-zero entries $R_{ij}$ are drawn following $\chi(d_{ij})$ distributions of $d_{ij}$ degrees of freedom:
\begin{equation}
\hspace{-0.2cm} \mathbf{R}=
\begin{bmatrix}
R_{1,1} &  & & \\
R_{2,1} & R_{2,2} & & \\ 
& R_{3,2} & R_{3,3} &\\ 
&  & \ddots & \ddots \\
\end{bmatrix}\sim\begin{bmatrix}
\chi(d_{1,1}) &  & & \\
\chi(d_{2,1}) & \chi(d_{2,2}) & & \\ 
& \chi(d_{3,2}) & \chi(d_{3,3}) &\\ 
&  & \ddots & \ddots \\
\end{bmatrix}.
\end{equation}
The random variable $Z_{ij}=\sqrt{\sum_{k=1}^{d_{ij}} X_k^2}$, where the $X_k\sim \mathcal{N}(0,1)$ are 
independent and identically distributed (i.i.d.), follows the $\chi(d_{ij})$ distribution. 
Equivalently, one can obtain a random variable that follows the $\chi(d_{ij})$ distribution by 
taking the square root of a random variable drawn from a chi-square distribution $\chi^2(d_{ij})$. 
The degrees of freedom $d_{ij}$ of the $\chi$ distributions in the $\beta$-Laguerre random matrix are:
\begin{align}
&\mathbf{d}=\begin{bmatrix}
n &  & & & \\
\beta(m-1) & n-\beta & & & \\
& \beta(m-2) & n-2\beta & & \\
&  &  \beta(m-3) & n-3\beta & \\ 
& & &  \ddots & \ddots   
\end{bmatrix} .
\end{align}
We recall that in this case $n=m\beta/\delta$ is  not  the dimension of the matrix $\mathbf{R}$, which is a $m\times m$ square matrix, but a positive real number.
Therefore, the rectangularity parameter does not need to be approximated in this method. 

Figure~\ref{fig:4s} shows numerical results of the random-matrix estimators of the average entropy-production minimum  for the 1D biased random walk with bias $A=1$. We draw $m\times m$ random matrices from the $\beta = 1,2,1000$ Laguerre ensembles. Note that $\beta = 1,2,4$ Laguerre ensembles are equivalent to Wishart random matrices with $\mathbf{R}_{ij}$ given respectively by real, complex and quaternionic normal random variables, and use the appropriate conjugate transpose of $\mathbf{R}$~\cite{dumitriu2002matrix}. 
We plot the maximum relative error $\epsilon_{\rm max}$~\eqref{Er} as a function of the size of the random matrix $m$ (Figure~\ref{fig:4s}a). The Wishart ($1$-Laguerre) ensemble provides a biased  overestimate of the real value with maximum relative error $2.3\%$ for small random matrices of sizes larger than  $64\times 64$. We observe that the $2$-Laguerre ensemble provides an estimator that is practically unbiased, even using small matrices (except in the limit $A\ll1$). Furthermore, $\beta$-Laguerre matrices with large values of $\beta$ (e.g. $\beta=1000$) yield small dispersion in the relative difference but a bias (underestimation) for small matrix sizes. The mean and the standard deviation of $\epsilon_{\rm max}$ obtained from a large population of computer-generated random matrices are observed to converge to zero with the matrix size $m$ as  $\sim 1/m$.  This fast convergence (compared to the usual $1/\sqrt{m}$) is a consequence of the correlation between the $m$ eigenvalues of the $\beta$-Laguerre random matrices. The convergence of the estimators is revealed in the difference between the spectral density of the random matrices and the Mar{\v{c}}enko-Pastur  distribution for $m=2^6$ (Fig.~\ref{fig:4s}b) and $m=2^{12}$ (Fig.~\ref{fig:4s}c). Remarkably, even though for $m=2^6$  the eigenvalue distribution is a rough approximation to the Mar{\v{c}}enko-Pastur distribution, the relative error of the estimator is smaller than $\pm$2.3\% for a single $1$-Laguerre and $\pm$1.5\% for a single $2$-Laguerre random matrix. 

 Eventually, we notice that all the expressions that involve a sum over the eigenvalues can be recast into random matrix traces. For instance, the two minimum entropy estimators and their associated maximum relative error read:
\begin{eqnarray}
\langle \tilde{S}_{\min}(t) \rangle_{k} &=& -Ak_- \;\int_0^t \tfrac{1}{m}\Tr e^{-M \bar{k}t'} \d t' \quad , \\
\langle \tilde{S}_{\min}(t) \rangle_{\tau}&=&\langle S_{\min}\rangle \; \left(1-\tfrac{1}{m}\Tr e^{-\mathbf{M}^{-1}t/\bar{\tau}} \right)\quad , \\
\epsilon_{\max}&=&(1-e^{-A})\tfrac{1}{m}\Tr \mathbf{M}^{-1} -1 \quad,
\end{eqnarray}
where $\mathbf{M}$ is either a Wishart random matrix or the Laguerre random matrix of size $m\times m$, scaled by its degree of freedom $n=e^A m$ (see Eqs.~(\ref{Wishart},\ref{Laguerre})).

\section{Explicit   expression for the average entropy production minimum}
\label{app:e}

We now employ  exact expressions for the moments of the Mar{\v{c}}enko-Pastur distribution $\rho$ to derive an analytical expression for the average minimum of the 1D biased random walk. The $n-$th moment ($n\geq 1$) of the Mar{\v{c}}enko-Pastur distribution $\rho(\lambda)$, can be expressed as a sum involving binomial coefficients:
\begin{align}
\left\langle \lambda^n\right \rangle_\rho=\sum_{r=0}^{n-1}\binom{n}{r}\binom{n-1}{r}\frac{\delta^r}{r+1}\quad .\label{eq:freemom}
\end{align}
 where $\langle ...\rangle_\rho$ denotes an average over the Mar{\v{c}}enko-Pastur distribution~\eqref{eq:MPdef}.
Then, it is convenient to expand the exponential in Eq.~\eqref{kform1}, and express the integrals  in terms of the moments~\eqref{eq:freemom}. By manipulating the indices, we obtain two infinite series:
 \begin{widetext}
\begin{align}
\langle S_{\min}(t)\rangle &= Ae^{-A}\sum_{n=1}^\infty \frac{(-k_+t)^n}{n!}\left\langle \lambda^{n-1}\right\rangle_{\rho} \\
&= Ae^{-A}\left(-k_+t+(k_+t)^2\sum_{r=0}^\infty\sum_{s=0}^\infty \frac{(-k_+t)^{r+s}}{(r+s+2)!}\binom{r+s+1}{r}\binom{r+s}{r}\frac{e^{-Ar}}{r+1}\right)  \quad.
\end{align}
\end{widetext}
This form can be identified as the \textit{Kamp\'e de F\'eriet} function $F$~\cite{exton1978handbook}, a two-variable generalization of hypergeometric functions. It is defined for integer vectors $\bf a,b,b',c,d,d'$, of lengths $p,q,q,r,s$ and $s$ respectively, as follows:
 \begin{widetext}
\begin{align}
{}^{p+q}F_{r+s}\left[\left.\genfrac{}{}{0pt}{}{\bf a\;, \;b\;, \; b'}{\bf c\;,\;d\;,\;d'} \right\rvert x\;,\; y \right]\equiv\sum_{m=0}^\infty\sum_{n=0}^\infty\frac{\prod\limits_{\alpha=1}^p (a_\alpha)_{m+n} \prod\limits_{\beta=1}^q (b_\beta)_{m}(b_\beta')_{n}}{\prod\limits_{\gamma=1}^r(c_\gamma)_{m+n} \prod\limits_{\delta=1}^s (d_\delta)_{m}(d_\delta')_{n}}\frac{x^my^n}{m!n!} \quad ,
\end{align}
\end{widetext}
where $(m)_n= \prod_{k=0}^{n-1}(m-k)=\frac{(m+n-1)!}{(m-1)!}$ denotes the rising factorial. The translations of the binomial coefficient into rising factorials yields the following expression of the mean minimum entropy:
\begin{widetext}
\begin{align}
\frac{\langle S_{\min}(t)\rangle}{A}&= -k_-t+{}^{2+0}\text{F}_{1+1}\left[\left.\genfrac{}{}{0pt}{}{ [2 \;1]\;, \;\varnothing\;, \; \varnothing}{ 3\;,\;2\;,\;2} \right\rvert -k_-t\;,\; -k_+t \right]\frac{k_-k_+t^2}{2} \quad .
\end{align}
\end{widetext}

\section{Sports data}

We provide additional details of the data used in Sec. VI about sports, selected from European competitions during the season 2018-2019.
For soccer, we use data from the UEFA Champions League in the group stage. We estimate the goal marking rate $k_+$ (goal conceding rate $k_-$) of the group leaders by summing the total number of goals scored (conceded) by all the group stage leaders divided by the cumulated duration of all the matches played in the group stage. There are 8 groups in Men's and 10 groups in Women's Champions League with 4 teams each. Each Men's team plays 6 matches, each Women's team plays 3 matches, each match lasts 90 minutes. Therefore, we select 48 matches of 8 teams out of 32 for Men's soccer and 30 matches of 10 teams out of 40 for Women's soccer.

For handball, we select the 2 best teams of each group (40 matches of 4 teams out of 12) in the main round of the Women's EHF Champions League, 2 best teams of each group (56 matches of 8 teams out of 28) in the group stage of the Men's EHF Champions League, each match lasts 60 minutes. For basketball, we select the 2 best teams of each group (56 matches of 4 teams out of 16) of the regular season of the Women's FIBA EuroLeague (the equivalent of the Champions League which does not exists for Women's basketball), and the best team or equally-ranked of each group (70 matches of 5 teams out of 32) of the regular season of the Men's FIBA Champions League, each match lasts 40 minutes.

%%%%


%%%%

\end{document}